\documentclass[journal=biochemistry,manuscript=article, layout=onecolumn]{achemso}
\setkeys{acs}{articletitle = true}
\usepackage[version=3]{mhchem} 
\usepackage[utf8]{inputenc} 
\usepackage{lineno}

\usepackage{xcolor}

\usepackage{xr}
\definecolor{red}{rgb}{0.75,0,0}
\definecolor{blue}{rgb}{0,0,0.75}
\definecolor{green}{rgb}{0,0.5,0}

\def\be{\begin{equation}}
\def\ee{\end{equation}}
\def\bea{\begin{eqnarray}}
\def\eea{\end{eqnarray}}

\def\besub{\begin{subequations}}
\def\eesub{\end{subequations}}

\def\bwd{\begin{widetext}}
\def\ewd{\end{widetext}}

\definecolor{MediumBlue}{RGB}{83,148,184}

\definecolor{MediumBlue}{RGB}{83,148,184}

\author{Anis Senoussi}
\affiliation[]{\small Sorbonne Université and CNRS, Laboratoire Jean Perrin, F-75005, Paris, France}

\author{Shunnichi Kashida}
\affiliation[]{\small Sorbonne Université and CNRS, Laboratoire Jean Perrin, F-75005, Paris, France}

\author{Ananyo Maitra}
\affiliation[]{\small Sorbonne Université and CNRS, Laboratoire Jean Perrin, F-75005, Paris, France}
\author{Raphael Voituriez}
\affiliation[]{\small Sorbonne Université and CNRS, Laboratoire Jean Perrin, F-75005, Paris, France}
\alsoaffiliation{Sorbonne Université and CNRS, Laboratoire de Physique Théorique de la Matière Condensée, F-75005, Paris, France}

\author{Jean-Christophe Galas}
\affiliation[]{\small Sorbonne Université and CNRS, Laboratoire Jean Perrin, F-75005, Paris, France}
\email{jean-christophe.galas@upmc.fr}

\author{André Estevez-Torres}
\affiliation[]{\small Sorbonne Université and CNRS, Laboratoire Jean Perrin, F-75005, Paris, France}
\email{andre.estevez-torres@upmc.fr}

\title[]{ 
\Large{Tunable corrugated patterns in an active gel sheet}
}

\begin{document}
\newpage

\begin{abstract}
Active matter locally converts chemical energy into mechanical work and, for this reason, it provides new mechanisms of pattern formation. In particular, active gels made of protein motors and filaments are far-from-equilibrium systems that exhibit spontaneous flow,\cite{Kruse2004, Voituriez2005} leading to active turbulence in two and three dimensions\cite{Sanchez2012, Kumar2018} and coherent flow in three dimensions\cite{Wu2017} (3D). 
Although these dynamic flows reveal a characteristic length scale resulting from the interplay between active forcing and passive restoring forces, the observation of static and long-range spatial patterns in active gels has remained elusive. In this work, we demonstrate that a 2D free-standing nematic active gel, formed spontaneously by depletion forces from a 3D solution of kinesin motors and microtubule filaments, actively buckles out-of-plane into a centimeter-sized periodic corrugated sheet that is stable for several days at low activity. Importantly, the corrugations are formed in the absence of flow and their wavelength and stability are controlled by the motor concentration, in agreement with a hydrodynamic theory. At higher activities these patterns  are transient with the gel becoming turbulent at longer times. Our results underline the importance of both passive and active forces in shaping active gels and indicate that a static material can be sculpted through an active mechanism.  
\end{abstract}


\newpage






Active matter is composed of subunits that convert free energy into mechanical work. It comprises systems composed of objects with very different sizes, from flocks of animals\cite{Cavagna2010} and bacterial colonies\cite{Dombrowski2004} to gels of cytoskeletal proteins\cite{Nedelec1997, Sanchez2012}. Active matter has attracted much attention, both from the theoretical and experimental perspectives, because it displays phase transitions and states that greatly differ from those observed at equilibrium, such as motile ordered states and spontaneous coherent or incoherent flow \cite{Toner2005,Julicher2007,Ramaswamy2010,Marchetti2013,Prost2015}. Among the active systems that can be studied in the laboratory, those composed of the protein filaments and motors that constitute the cytoskeleton of the eukaryotic cell are of special interest for three reasons: i) their biological importance\cite{Dogterom2013, Blanchoin2014}, ii) the possibility to make purified systems that can be easily controlled and studied\cite{Nedelec1997, Sanchez2012, Bendix2008}, and iii) their potential to make useful materials\cite{Needleman2017}. 

Depending on the conditions, cytoskeletal active systems display a wide array of dynamic behaviors. When in contact with a solid surface, they generate polar patterns\cite{Schaller2010} and large scale vortices\cite{Sumino2012} in two dimensions (2D). 
In contrast, when fluid interactions strongly influence the dynamics, as in active gels, asters and vortices of filaments have been observed\cite{Nedelec1997} in 3D motor-microtubule assays at low densities. At higher concentrations, active turbulence has been reported both in 2D\cite{Sanchez2012, Doostmohammadi2018, Kumar2018} and 3D\cite{Sanchez2012, Wu2017} and global contraction\cite{Bendix2008, Alvarado2013, Foster2015, Torisawa2016} and spontaneous flow\cite{Wu2017} have been  reported in 3D.
Although these behaviors may exhibit a characteristic length scale, resulting from the combination of active forcing and passive restoring forces, the observation of static spatial patterns in dense active gels has remained elusive. 

Here we report the observation of a novel behavior in an active gel: the formation of a thin corrugated sheet in three dimensions. The gel is prepared inside a shallow and long channel of rectangular cross-section with nematic order along the longest axis of the channel. The gel contracts anisotropically along its two shortest dimensions to form a thin sheet of gel that freely floats in the aqueous solution, mainly due to passive depletion forces. Simultaneously, the extensile active stress generated by the motors buckles the sheet along the direction perpendicular to its plane, forming a corrugated sheet of filaments with a well-controlled  wavelength of the order of 100~$\mu$m that spans an area of 10~mm$^2$. This out-of-plane buckling occurs in the absence of hydrodynamic flow within the gel, and is distinct from the flow-generating in-plane buckling that is common in active gels.\cite{Kruse2004, Voituriez2005, Sanchez2012, Wu2017} The corrugated patterns are either stable, at low activity, or break into an active turbulent state at high activity. We find that the wavelength of the patterns scales with the inverse of the motor concentration, which can be accounted for with a hydrodynamic theory and shows that an active mechanism can be used to shape static materials.


\begin{figure}[h!]
	\centering
	\includegraphics[width=\linewidth]{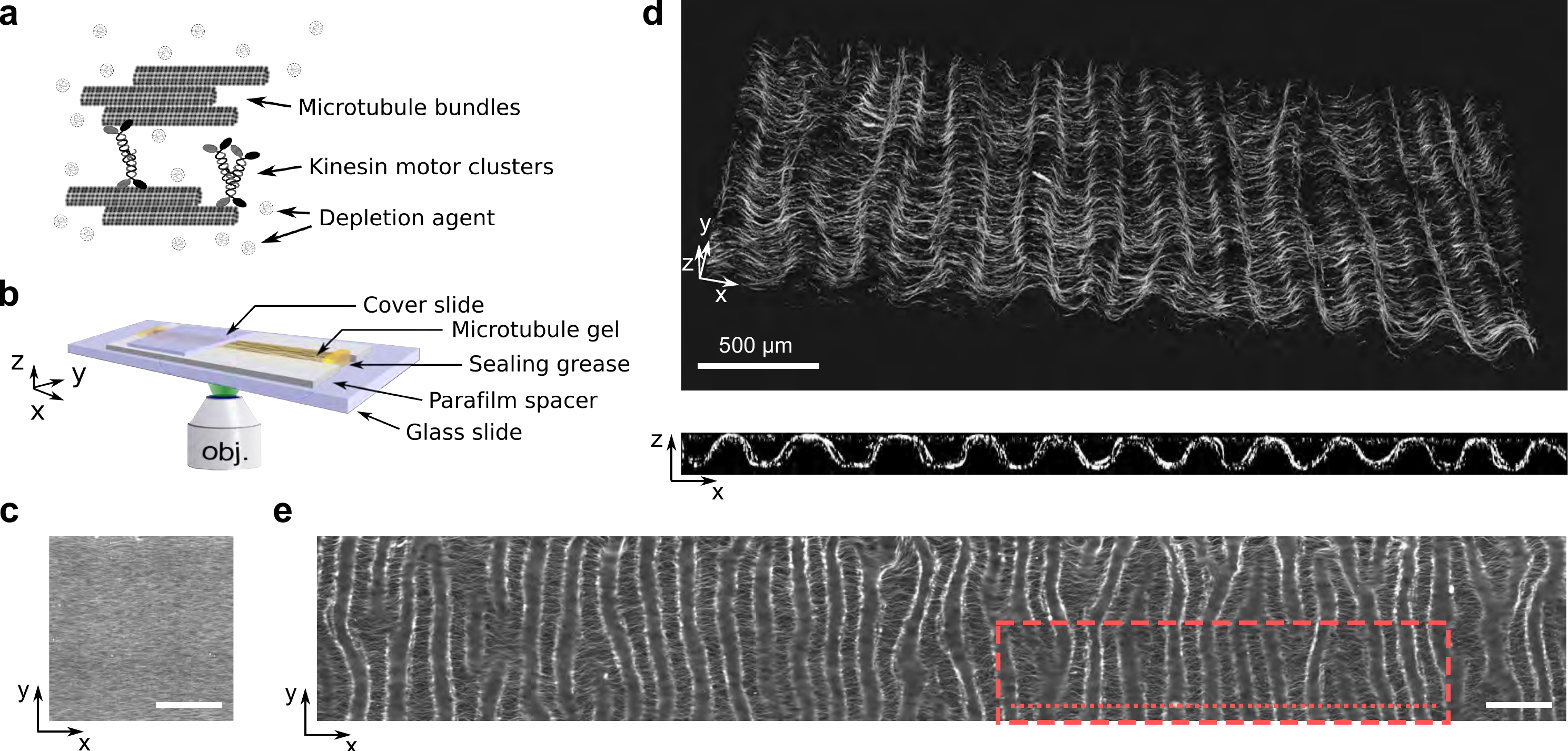}
	\caption{At low motor concentration a 3D active nematic gel creates a  thin corrugated sheet of well defined wavelength. \textbf{a} Scheme of the components of the active gel formed by non-growing microtubules bundled together by a depletion agent and clusters of kinesin motors. \textbf{b} Scheme of the channel where the gel (in yellow) is observed.  \textbf{c} Epifluorescence image of the gel at initial time. \textbf{d} Confocal images in 3D (top) and cross-section in the $xz$ plane (bottom) of the gel after 300 min. \textbf{e} Epi-fluorescence image of the same sample after one day and over a $9.5\times1.4$~mm$^2$ area, the red dashed rectangle and the red dotted line respectively  indicate the region where the top and bottom images in panel d were recorded. Scale bars are 500~$\mu$m and motor concentration 0.5 nM.\label{Fig_1}} 
\end{figure}

\begin{figure}[htbp]
	\centering
	\includegraphics[width=0.5\linewidth]{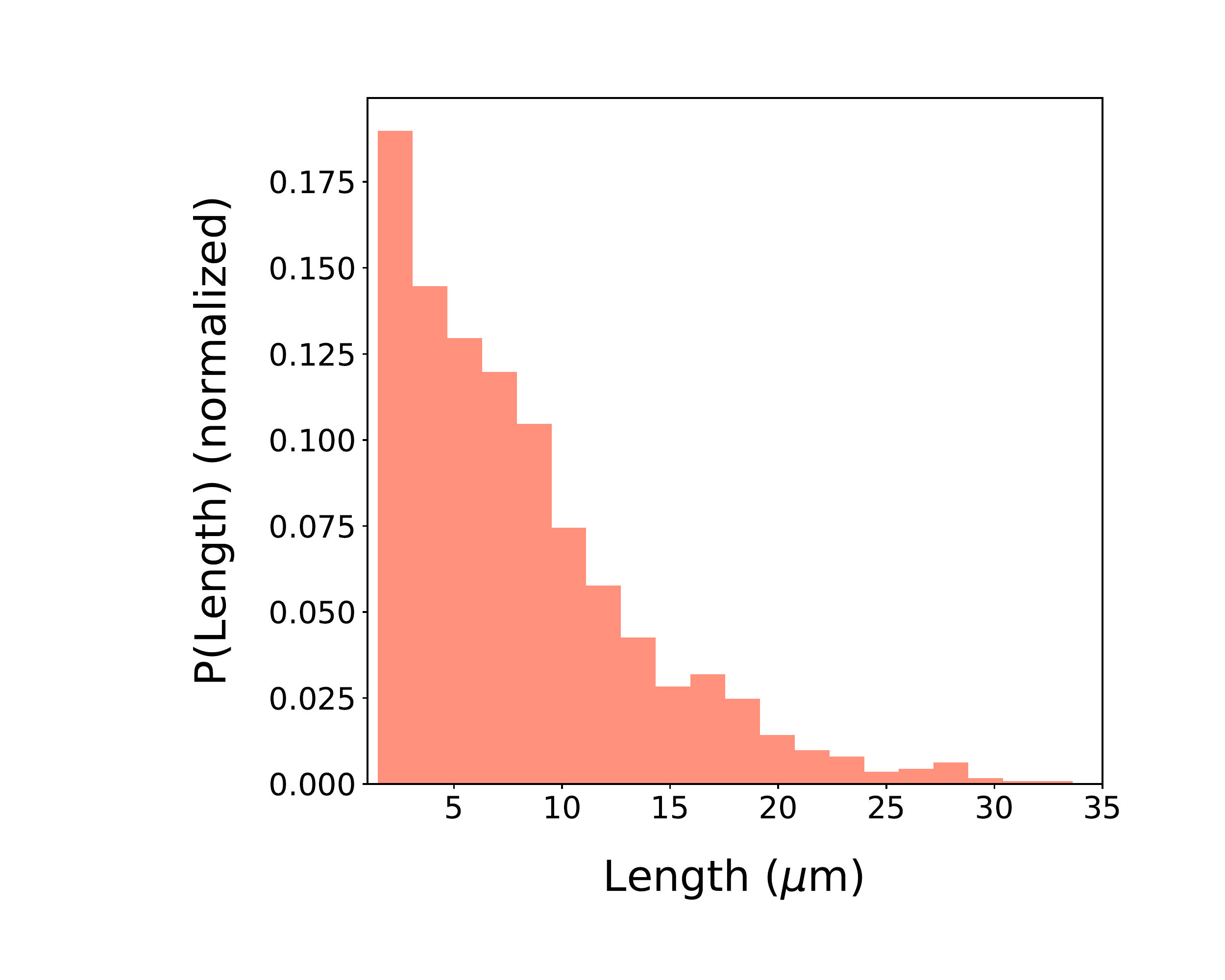}
	\renewcommand{\figurename}{Figure S\!\!}\caption{Normalized probability distribution of taxol-stabilized microtubules in our experiments. Microtubules have an average length of 8.1 $\mu$m with a standard deviation of 5.8 $\mu$m.}\label{Fig_SI_plength} 
\end{figure}

The gel is constituted of a dense suspension of non-growing microtubules bundled together by a depletion agent and by clusters of kinesin-1 motors (Figure~\ref{Fig_1}a). It is supplemented with ATP and an ATP-regeneration system that keeps the motor active for at least 4 h. This is similar to previously published active nematic gels\cite{Sanchez2012} but it differs in several important ways: the microtubules are longer ($8\pm6~\mu$m instead of 1~$\mu$m, Figure~S\ref{Fig_SI_plength}), the kinesin used here\cite{furuta2013}, K430, is different from the standard K401 (it comes from a different organism and forms non-specific clusters), and its typical concentration is two orders of magnitude lower (see SI Section \ref{SI_methods}). In the following we find that the motor concentration and, specially, the microtubule length, are key to explain our observations. 

\begin{figure}[htbp]
	\centering
	\includegraphics[width=\linewidth]{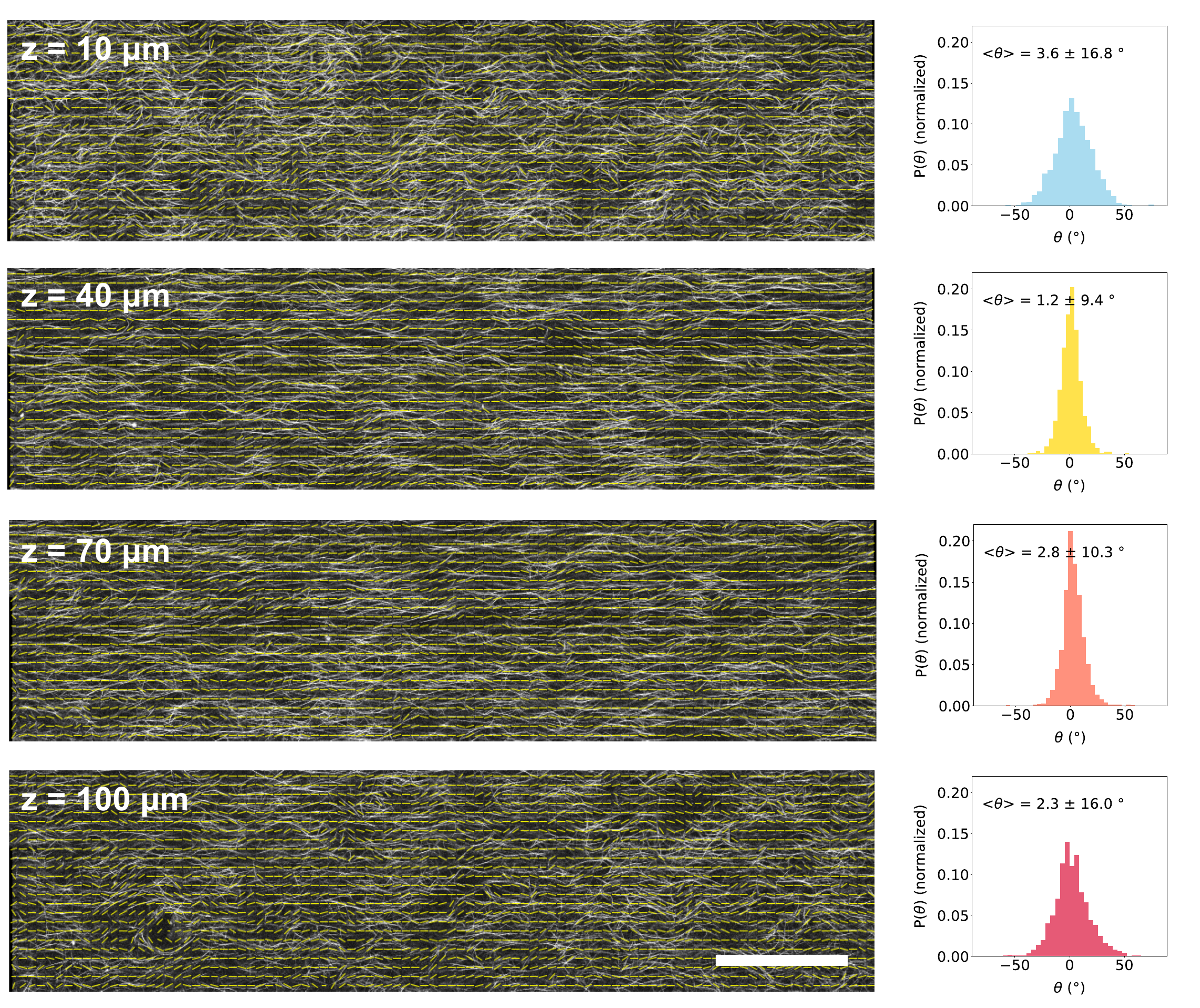}
	\renewcommand{\figurename}{Figure S\!\!}\caption{Initial orientation of microtubules in the $xy$ plane for different heights in a flow cell. Orientational fields and distributions were obtained using OrientationJ plugin\cite{orientationJ} for ImageJ. $\theta$ is the angle between the filaments and the $x$ axis in the $xy$ plane.  Scale bar 250 $\mu$m.} \label{Fig_SI_orientation_0min} 
\end{figure}

The gel is prepared in a rectangular channel with length $L=22$~mm, width $W=1.5$~mm and height $H=0.13$~mm (see SI Methods), with $L\gg W \gg H$, such that the microtubule bundles are aligned along $L$, parallel to the $x$ axis, and the height of the channel  is parallel to the $z$ axis (Figure~\ref{Fig_1}b-c). This initial nematic order arises spontaneously during the filling process of the channel by capillarity and the angle of the director of the nematic with the $x$ axis is $2\pm16$\textsuperscript{o} (Figure S\ref{Fig_SI_orientation_0min}).  The microtubule bundles are fluorescent because they bear a small fraction of fluorescent tubulin, allowing the observation of the gel by fluorescence microscopy. 
In the presence of 0.5 nM of motors, confocal images recorded after 300~min show that the gel has buckled in the $xz$ plane to form a corrugated sheet whose hills and valleys reach the top and bottom walls of the channel and whose grooves are strikingly parallel to the $y$ axis (Figure~\ref{Fig_1}d). The thickness of the sheet is  $\ell_z= 35\pm5$~$\mu$m and the wavelength of the corrugations is $\lambda=285\pm15~\mu$m. This periodic pattern extends along an area of at least $9.5\times1.4$~mm$^2$, with some defects corresponding to the junction of two valleys or hills. The pattern can also be visualized in epifluorescence, where it appears in the form of focused and defocused bands (Figure~\ref{Fig_1}e). 
\begin{figure}[h!]
	\centering
	\includegraphics[width=\linewidth]{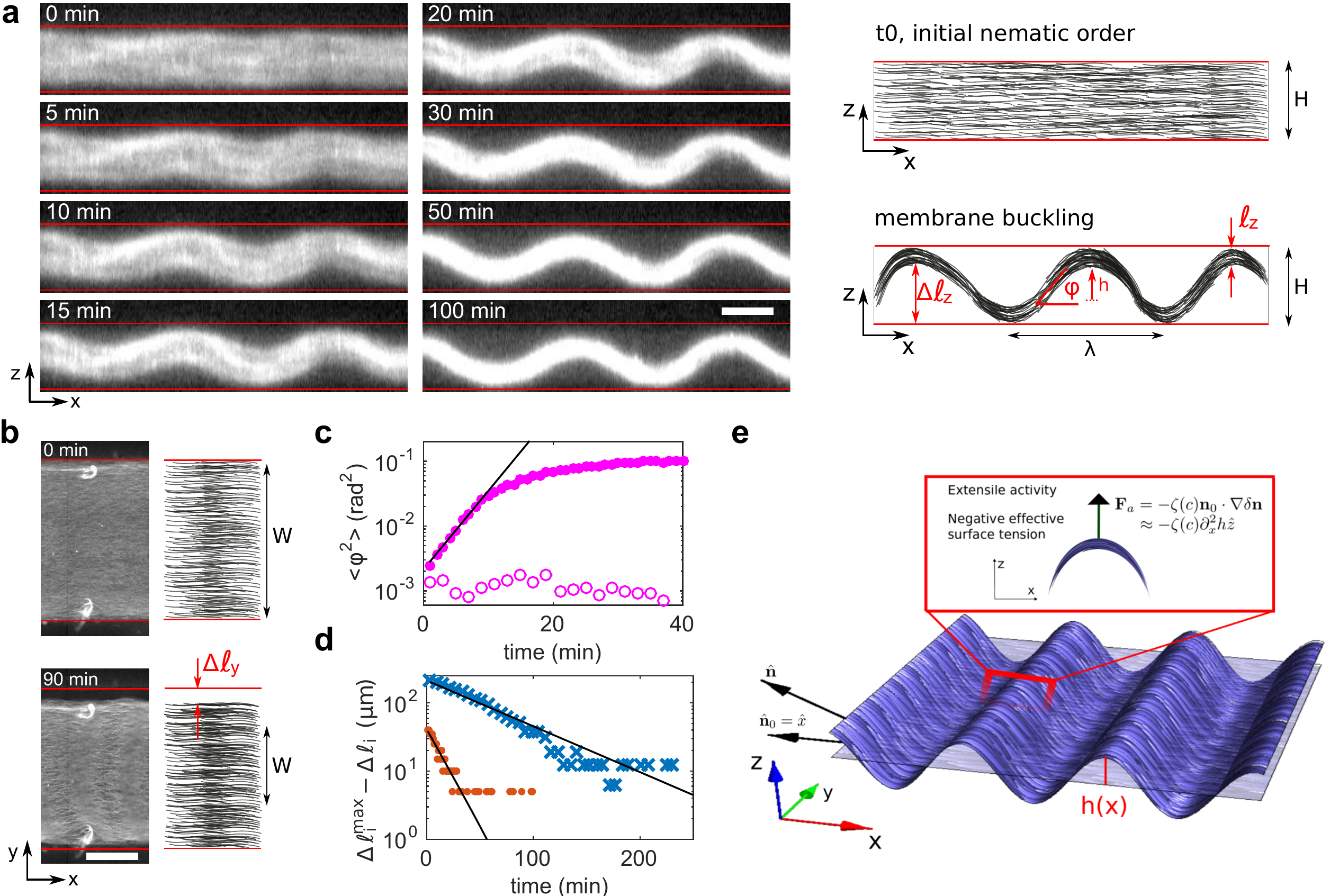}
	\caption{Dynamics and mechanism of the formation of a thin corrugated sheet at low motor concentration. \textbf{a} Time-lapse confocal fluorescence images of the active gel in the $xz$ plane (left) and sketch of the observations indicating the measured quantities $\Delta\ell_z$, $\phi$, $h$,  and $\ell_z$. Scale bar is 100 $\mu$m. \textbf{b} Epifluorescence images of the gel at $t=0$ and 90~min (left) and sketch indicating the measured quantity $\Delta\ell_y$. Scale bar is $500~\mu$m. Red lines in panels a and b indicate channel walls.
\textbf{c}  Average of $\phi^2$ along the $x$ direction \emph{vs.} time in the presence (filled disks) and in the absence of motors (empty circles). \textbf{d} Offset to the maximum contracted length along the $z$ (red disks) and $y$ (blue crosses) directions. Black lines in panels c and d are exponential fits. \textbf{e} Sketch of the mechanism for the active buckling of a thin membrane through the negative tension $F_a$ proportional to the active stress $\zeta(c)$ and the Laplacian of the height $h(x)$ of the sheet above its fiducial plane. All data correspond to 0.5 nM motors except empty circles in panel c.}
\label{Fig_2} 
\end{figure}

To elucidate the mechanism of pattern formation we recorded confocal (video S\ref{Video_SI_video_fig2MT}) and epifluorescence time-lapse images of a buckling gel at 0.5 nM motors (Figure~\ref{Fig_2}). When we look at the corresponding $xz$ and $xy$ views of the gel (Figure~\ref{Fig_2}a-b) three processes can be identified: buckling along the $z$ direction and contractions along $z$ and $y$, that we quantify by the angle $\phi$ between the microtubule bundles and the $x$ axis in the $xz$ plane, and by $\Delta\ell_z$ and $\Delta\ell_y$, the contracted lengths of the gel along the $z$ and $y$ axes. To quantify the initial rate of buckling we measured the growth over time of $\langle\phi^2\rangle$, the average of $\phi^2$ along the field of view in $x$ of the confocal image, and obtained $\omega_{\phi}=0.3$~min$^{-1}$. Buckling later proceeded at a slower pace until reaching a maximal buckling angle $\phi^{max} = 32.2\pm0.5$\textsuperscript{o} (Figure~\ref{Fig_2}c) and amplitude $h^{max} = 22\pm3$~$\mu$m after 100 min. Contraction along the $z$ and $y$ directions was significantly slower with onset rates $\omega_z=6.4\times10^{-2}$~min$^{-1}$ and $\omega_{y}=1.5\times10^{-2}$~min$^{-1}$, respectively, to reach maximum amplitudes $\Delta \ell_z^{max} = 40$~$\mu$m and $\Delta \ell_y^{max} = 210$~$\mu$m (Figure~\ref{Fig_2}d). Note that the relative contraction amplitudes $\Delta \ell_z^{max}/H = 0.40$ 
 and $\Delta \ell_y^{max}/W = 0.14$ 
 are significantly different, indicating that the final contracted state does not correspond to a nematic liquid droplet at equilibrium\cite{deGennes1993}.  

\begin{figure}[htbp]
	\centering
	\includegraphics[width=\linewidth]{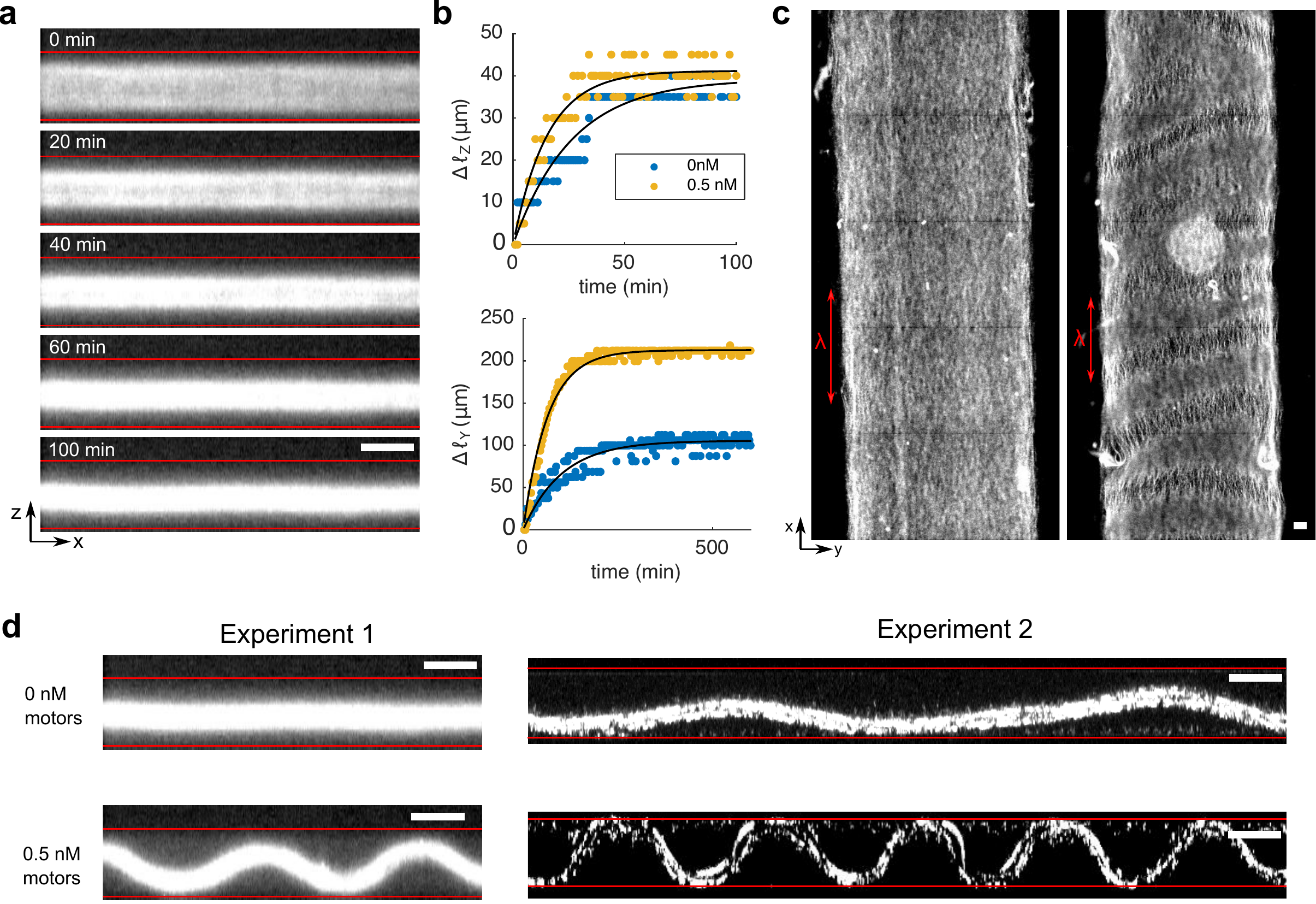}
	\renewcommand{\figurename}{Figure S\!\!}\caption{Passive formation of the gel sheet and comparison of the buckling in the passive and active case, including reproducibility. a. Time-lapse confocal fluorescence images of the passive gel in the $xz$ plane. b. Comparison of contractions along $z$ (top) and $y$ (bottom) between gels with and without motors. c. Epifluorescence images in the absence (left) and in the presence (right) of motors show focused and defocused bands resulting from buckling with $\lambda$ indicating the wavelength. The pattern in the absence of motors is faint. d. Buckling reproducibility for experiments performed in a 3 weeks interval with different batches of microtubules. The extent of buckling in the absence of motors is variable (top) while it is fairly reproducible in its presence (bottom). Scale bars are 100 $\mu$m.}\label{Fig_SI_compare_kin_no_kin} 
\end{figure}

\begin{figure}[htbp]
	\centering
	\includegraphics[width=0.4\linewidth]{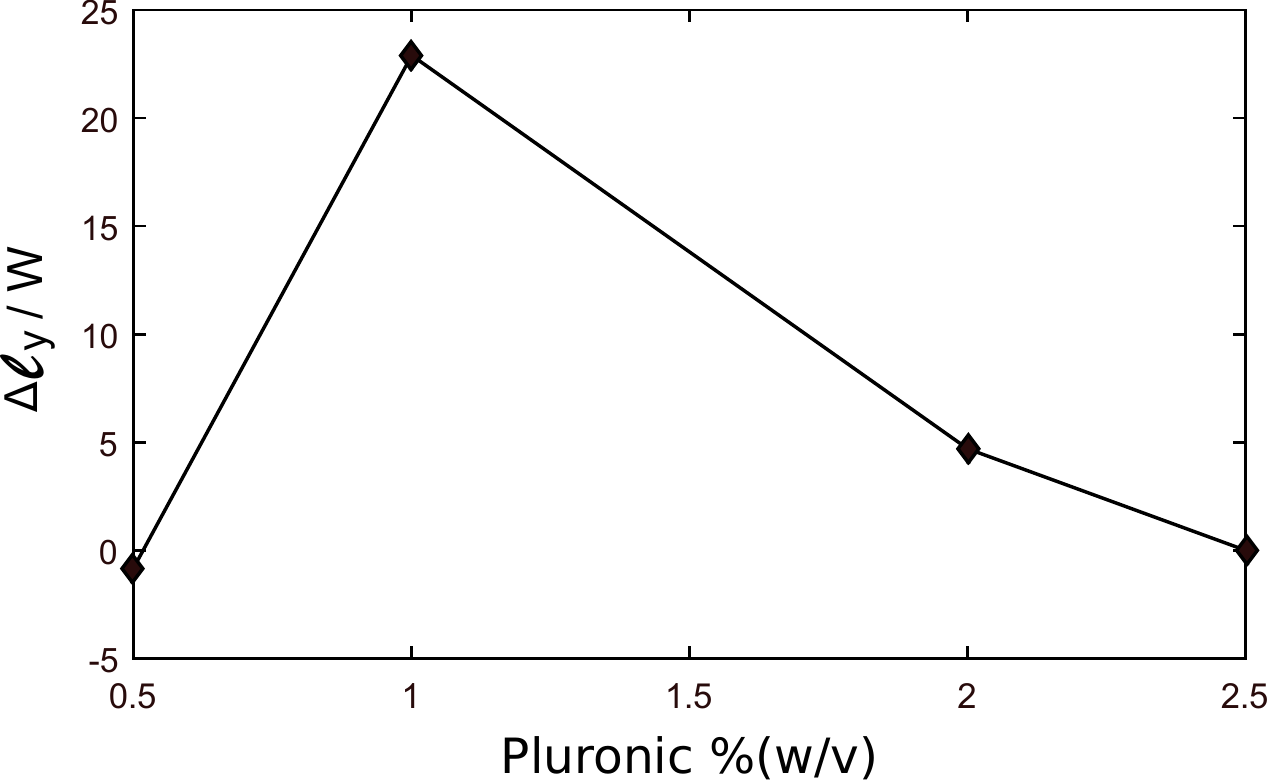}
	\renewcommand{\figurename}{Figure S\!\!}\caption{Gel contraction along the $y$-axis as a function of pluronic concentration. Contraction depends non-monotonically on the concentration of the depletion agent, with a maximum at 1 \%(w/v) pluronic concentration.}\label{Fig_SI_pluronic_range_no_motors} 
\end{figure}

\begin{figure}[H]
	\begin{center}
		\includegraphics[width=0.75\linewidth]{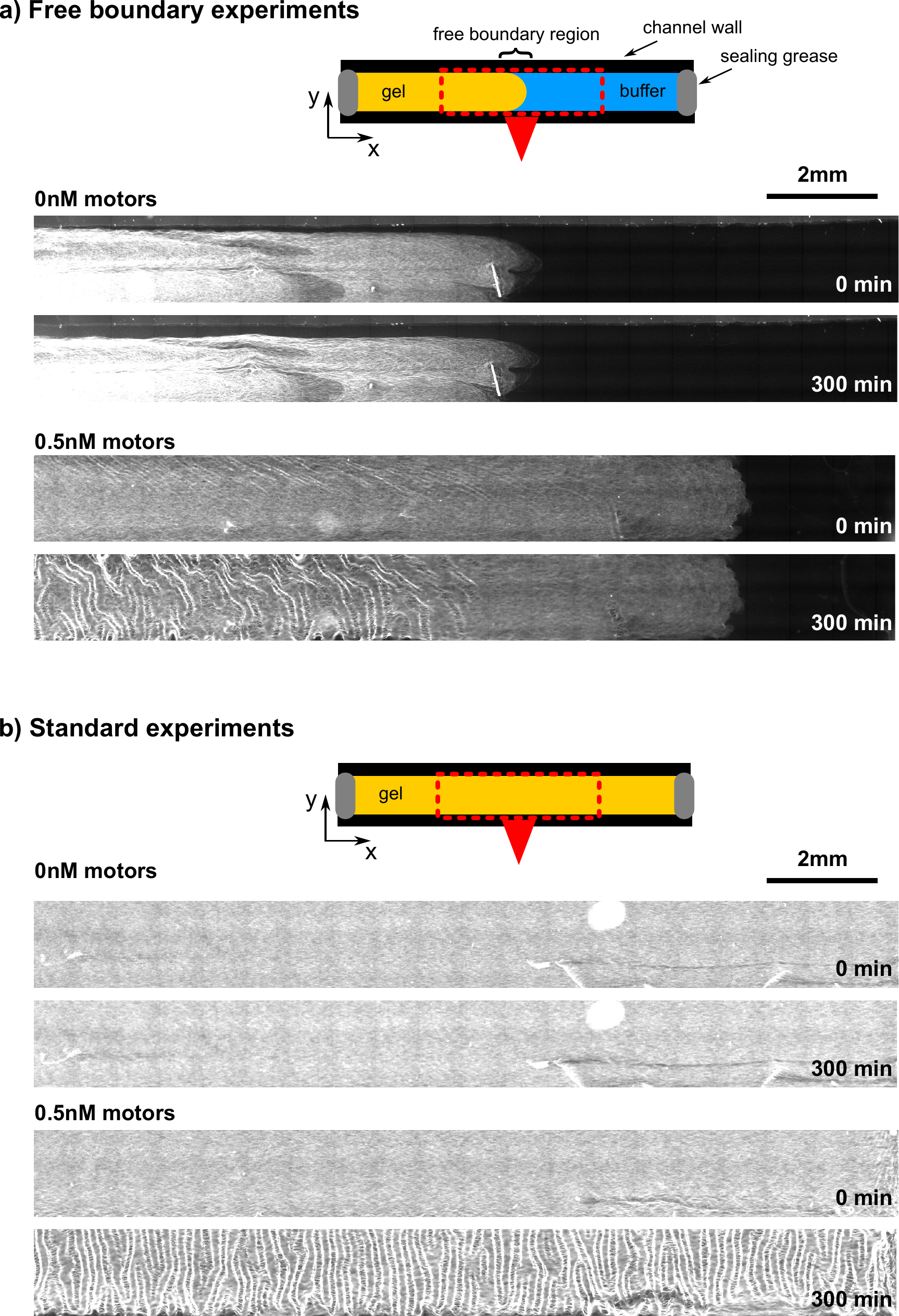}
		\renewcommand{\figurename}{Figure S\!\!}\caption{Behavior of passive and active gels with and without a free boundary. \textbf{a} Scheme showing a top view of a gel with a free boundary on the right side and a solid boundary on the left side. Dashed red squares indicate regions where images where recorded. Fluorescent images corresponding to these regions at different times in the absence and in the presence of motors. The white oblique line on the right side of the passive gel is an impurity. \textbf{b} Similar experiments with boundaries on both sides. The corrugated pattern forms in the presence of motors both with and without boundaries. In contrast, passive gels with free boundaries do not form corrugations.} \label{Fig_SI_boundary}
	\end{center}
\end{figure}

Control experiments in the absence of motors (Figures~\ref{Fig_2}c and S\ref{Fig_SI_compare_kin_no_kin}) 
show that passive gels contract similarly to active ones but they buckle significantly less. $\Delta \ell_z^{max}=38$~$\mu$m was comparable with the active gel while $\Delta \ell_y^{max}= 105~\mu$m and the rates $\omega_z=3.8\times10^{-2}$~min$^{-1}$ and $\omega_{y}=1.0\times10^{-2}$~min$^{-1}$ were two-fold smaller. In addition, $\Delta \ell_y^{max}$ in the absence of motors depends on the concentration of the depletion agent, here pluronic, indicating that contraction is due to depletion forces (Figure~S\ref{Fig_SI_pluronic_range_no_motors}). In confocal images with the same field of view in $x$ as above (660 $\mu$m), buckling is undetectable with $\langle\phi^2\rangle$ remaining constant at $10^{-3}$~rad$^2$ (Figure~\ref{Fig_2}c), $\phi^{max} = 5\pm2$\textsuperscript{o} and $h^{max} = 1\pm1$~$\mu$m after 100 min, all significantly smaller than the values obtained in the presence of motors. However, passive buckling is weak but distinguishable in epifluorescence images acquired over a wider field of view, although the pattern shows more variability (Figure~S\ref{Fig_SI_compare_kin_no_kin}). In addition, passive buckling is only observed when the gel is constrained between two hard boundaries in the $x$ direction, while active buckling is observed both in the presence and in the absence of such hard boundaries (Figure~S\ref{Fig_SI_boundary}). Taken together, these experiments demonstrate, firstly, that passive and active buckling happen through different mechanisms and, secondly, that in active gels buckling is principally an active mechanism while contractions in $y$ and $z$ are mainly passive. In passive gels, depletion forces induce the condensation of microtubules into a dense nematic gel phase, which, in the absence of confinement, would relax to a highly anisotropic tactoid droplet\cite{Kaznacheev2002}. In the geometry of our experiments, this results in the formation of a quasi 2D sheet that elongates along the nematic axis $x$, thereby leading to Euler buckling in the presence of boundaries. When the gel is active, although depletion forces are still crucial to make a thin sheet, the activity of motors controls its buckling and determines its wavelength even in the absence of boundaries, as we argue below.

We constructed a hydrodynamic theory to model the periodic undulation of the thin microtubule sheet in the $xz$ plane and to obtain a theoretical estimate of the wavenumber $q^*=2\pi/\lambda$ of the pattern (see Figure~\ref{Fig_2}e and SI Section \ref{SI_theory}). We consider the microtubule-kinesin system to be a thin nematic active gel with bending modulus $K$ and the director $\hat{{\bf n}}$ on average being parallel to the $x$ axis: $\hat{{\bf n}}_0=\hat{x}$. The fluctuation of the membrane about a fiducial plane parallel to the $xy$ plane (here, taken to be the mid-plane of the channel) is denoted by $h(x,y)$.
The deflection of the director in the $xz$ plane, $\delta{\bf n}_z$, leads to a buckling of the membrane in the $z$ direction: $\delta{\bf n}_z\approx \partial_x h\hat{z}$.
The passive elasticity of the nematic gel $\propto (\nabla {\bf n})^2$ then yields a bending energy $\propto (K/2)(\partial_x^2h)^2$ for the buckling of the thin sheet in the $z$ direction. The standard active force\cite{Toner2005,Julicher2007,Ramaswamy2010,Marchetti2013,Prost2015} is $-\zeta(c)\nabla\cdot({\bf nn})$, where $\zeta(c)>0$ is the strength of the extensile activity that is a function of motor concentration $c$. This leads to a force $\propto -\zeta(c)\partial_x^2h\hat{z}$ that tends to destabilise the flat membrane and that is similar to an effective \emph{negative} surface tension \cite{Maitra2014}. The interplay between the negative surface tension, arising from activity, and the stabilizing bending modulus, due to nematic elasticity, leads to the selection of a pattern with wavenumber $q^*\sim\sqrt{\zeta(c)/K}\sim c/\sqrt{K}$, where we used the fact that the activity coefficient scales as the square of the motor concentration\cite{Martinez-Prat2019}. Although this active buckling instability results from the interplay of active forcing and passive elastic restoring forces, as in other active gel experiments \cite{Sanchez2012, Wensink2012, Keber2014, Zhou2014, Kumar2014, DeCamp2015, Duclos2016, Duclos2018, Opathalage2019, Martinez-Prat2019}, here the out-of-plane buckling of the active sheet precedes any planar pattern formation, in contrast to those experiments. In addition, the instability described here does not result in coherent or incoherent flow, of either the gel or the embedding fluid, in contrast with theories describing 2D or 3D active gels that do not form sheets\cite{Kruse2004, Voituriez2005, Edwards2009}. Furthermore, the three-dimensional conformation of the buckled membrane is remarkably stable and static at small motor concentrations, in contrast to other active gel experiments that quickly evolve to a dynamic, coherently or incoherently, flowing state.

\begin{figure}[h!]
	\centering
	\includegraphics[width=\linewidth]{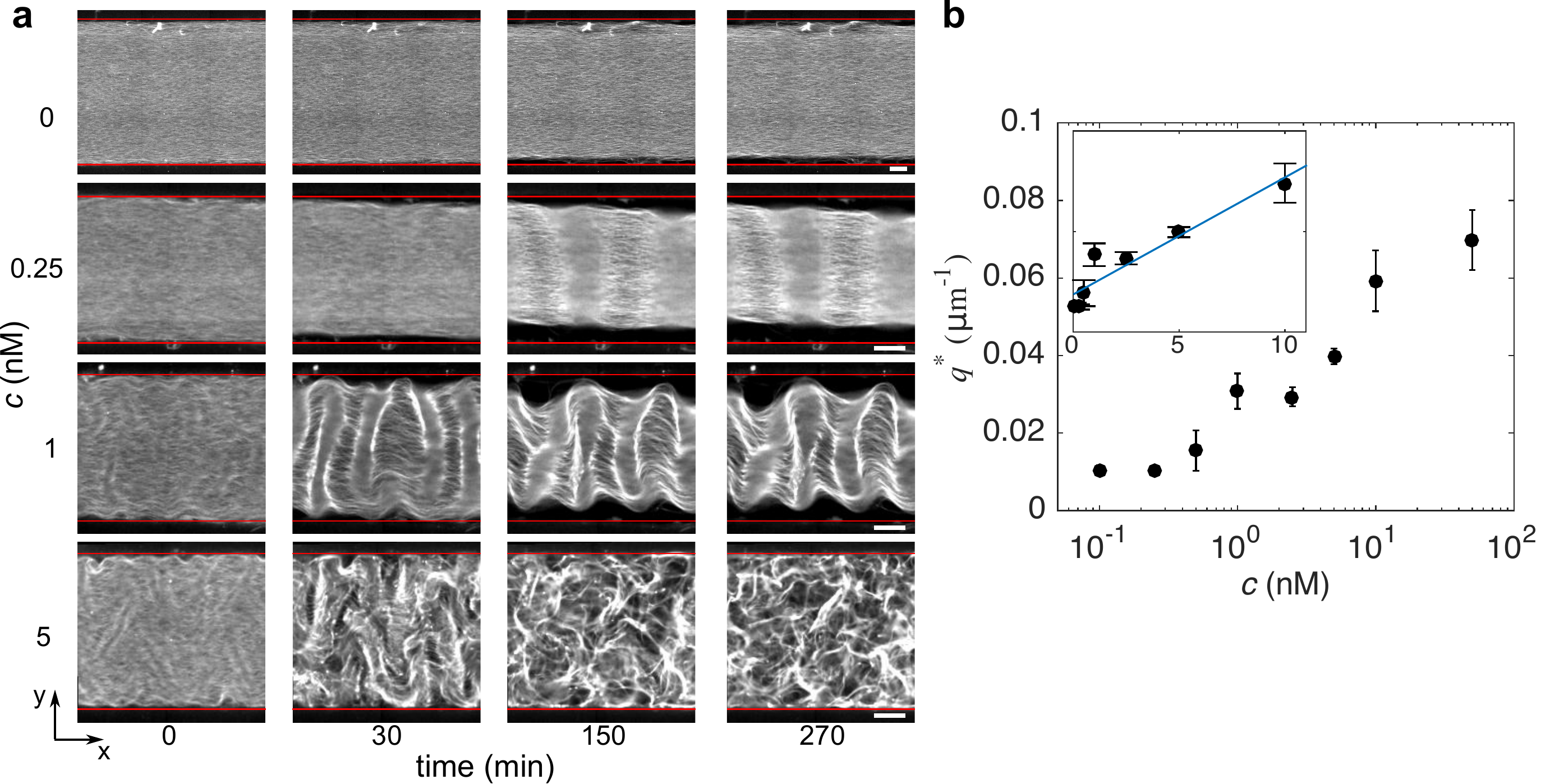}
	\caption{Dynamics, shape and stability of the patterns strongly depend on motor concentration. \textbf{a} Time-lapse epifluorescence images of gels with different motor concentrations $c$. Red lines indicate channel walls. Scale bars are 200 $\mu$m. \textbf{b} Wavenumber $q^*$ of the corrugations \emph{vs.} motor concentration. The inset is the lin-lin representation of the main plot and the blue line corresponds to a linear fit to the data in the range $c=0.1-10$~nM. Error bars indicate the standard deviation of a triplicate experiment. \label{Fig_3}} 
\end{figure}

\begin{figure}[h!]
	\centering
	\includegraphics[width=\linewidth]{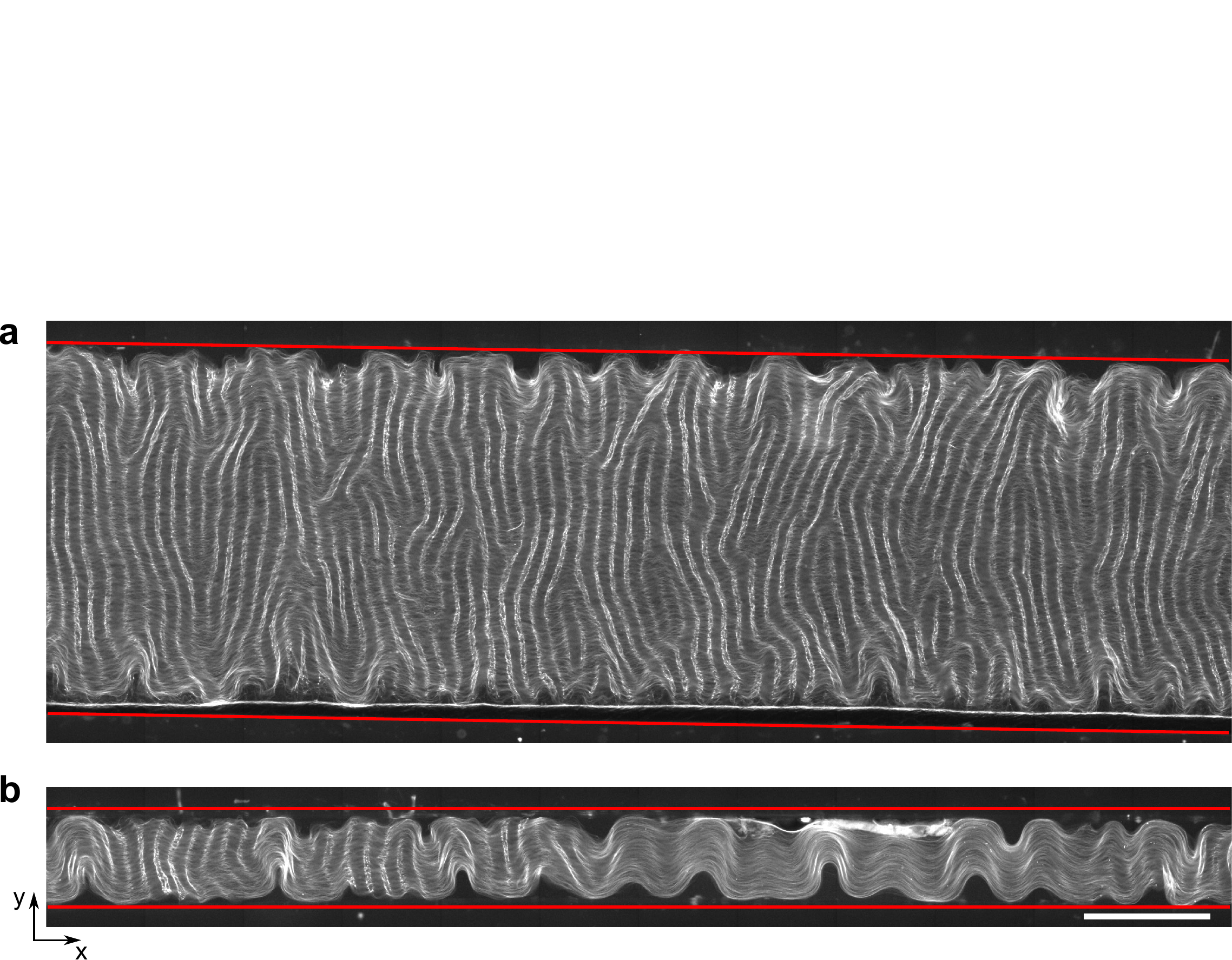}
	\renewcommand{\figurename}{Figure S\!\!}\caption{The pattern depends on channel width. Fluorescence image of the gel after 39 min with 0.5 nM motors for channels of different width: 2.9 (\textbf{a}) and 0.6 mm (\textbf{b}), and the same height (0.13 mm). Red lines indicate channel walls. Scale bar is 1 mm.}\label{Fig_SI_varying_width} 
\end{figure}

To test our theory we investigated the behavior of the gel over a range of motor concentrations $c$ spanning more than two orders of magnitude (Figure~\ref{Fig_3} and video S\ref{Video_SI_epifluo_kin_range}). Below 0.5 nM motors, the gel behaves as described in Figure~\ref{Fig_2}:  buckling in the $xz$ plane and contractions in the $z$ and $y$ directions. In agreement with the theory (SI Section \ref{SI_theory}), when the width $W$ of the channel is reduced such that $W\sim H$, some portions of the gel buckle in $xz$ and others in the $xy$ plane (Figure~S\ref{Fig_SI_varying_width}). As $c$ increases, between 1 and 2.5 nM motors, buckling in the $xz$ plane is initially observed and followed by buckling in the $xy$ plane that distorts the corrugated pattern without breaking it. In all the cases where both $xy$ and $xz$ buckling is observed, the wavelength of the former is longer than that of the later, in agreement with the theory that predicts $\lambda\sim\sqrt{K}$, as we expect $K$ to scale with the thickness of the gel that is larger along $y$ than along $z$. Finally, between 5 and 50 nM motors, buckling in the $xz$ plane is still observed at early times but the pattern breaks into a 3D active turbulent state similar to the one already reported in this type of gel\cite{Sanchez2012, Wu2017} (videos S\ref{Video_SI_video_BandsToTurbulence} and S\ref{Video_SI_confocal_bands_to_turbulence}). However, the velocity of this flow state is significantly lower in our case, possibly because the gel is more viscous.  We interpret this transition to turbulence with the same instability as previously proposed\cite{Martinez-Prat2019} but from a different starting point: a corrugated sheet instead of 2D or 3D nematic gel. Importantly, the measured wavenumber of the corrugations is in good agreement with the predicted linear scaling (Figure~\ref{Fig_3}b), in particular in the range 0.5-10~nM. A linear fit $q^*=a_1+a_2c$ of the data yields $a_1= 5\times10^{-3}$~$\mu$m$^{-1}$ and $a_2= 1.4\times10^{-3}$$\mu$m$^{-1}$nM$^{-1}$, where the constant term $a_1$ results from the passive mechanism of Euler buckling.  

\begin{figure}[htbp]
	\centering
	\includegraphics[width=1\linewidth]{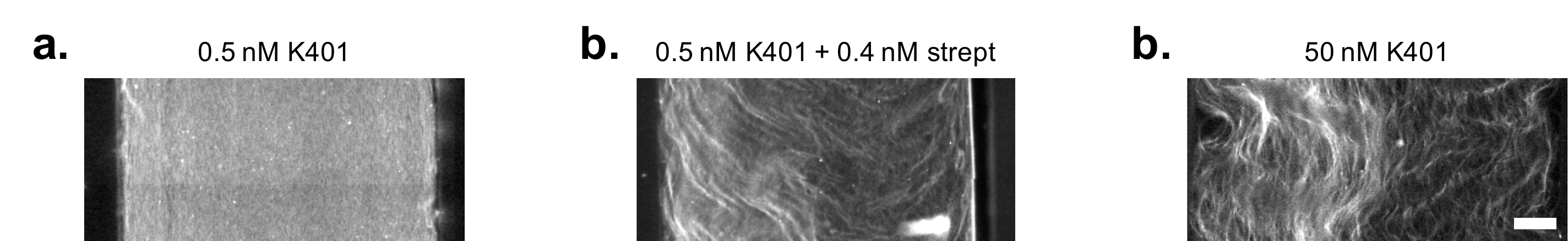}
	\renewcommand{\figurename}{Figure S\!\!}\caption{Corrugated patterns are observed with the standard motor K401. a. No patterns were obtained using K401 - that bears a biotin that binds to tetrameric streptavidin forming K401 clusters - at 0.5 nM without streptavidin. b. The addtion of 0.4 nM of streptavidin led to patterns. c. K401 formed periodic patterns at 5 nM in the absence of streptavidin, suggesting that 1 \% of K401 make non-specific clusters at these concentrations. Scale bar 250 $\mu$m.}\label{Fig_SI_K401BCCP} 
\end{figure}

\begin{figure}[htbp]
	\centering
	\includegraphics[width=0.8\linewidth]{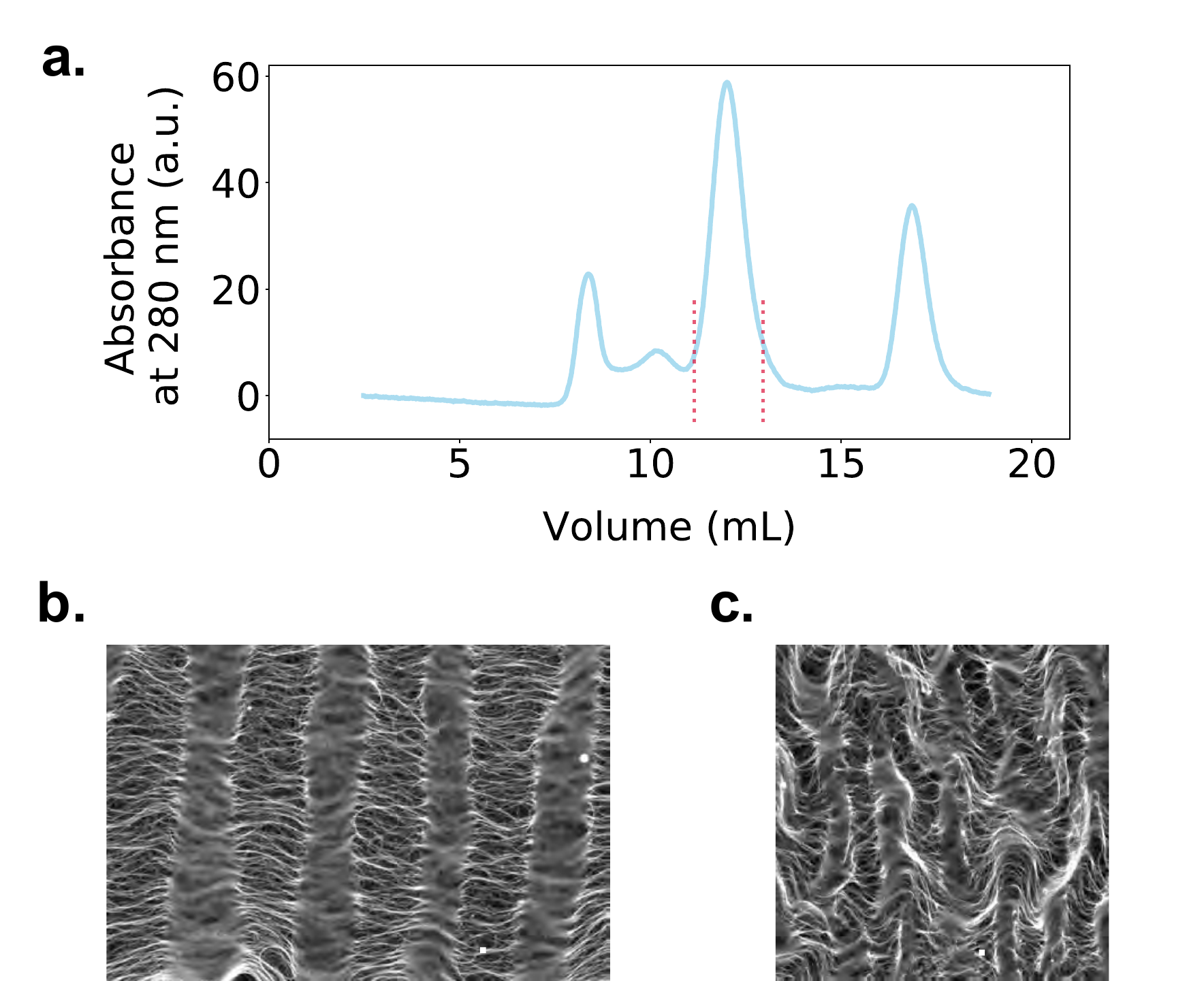}
	\renewcommand{\figurename}{Figure S\!\!}\caption{a. Gel filtration profile of dialyzed kinesin. K430 dimers come out at 12 mL (between red dashed lines). Peak at 17 mL corresponds to the SNAP-tag which was produced alone. b. Periodic pattern obtained with 0.22 nM of proteins which were not gel purified (fractions between 7.5 mL and 13 mL). c. Periodic pattern obtained with 14 nM of protein aggregates (peak 8.5 mL). Scale bars 250 $\mu$m.}\label{Fig_SI_akta} 
\end{figure}

\begin{figure}[h!]
	\begin{center}
		\includegraphics[width=0.8\linewidth]{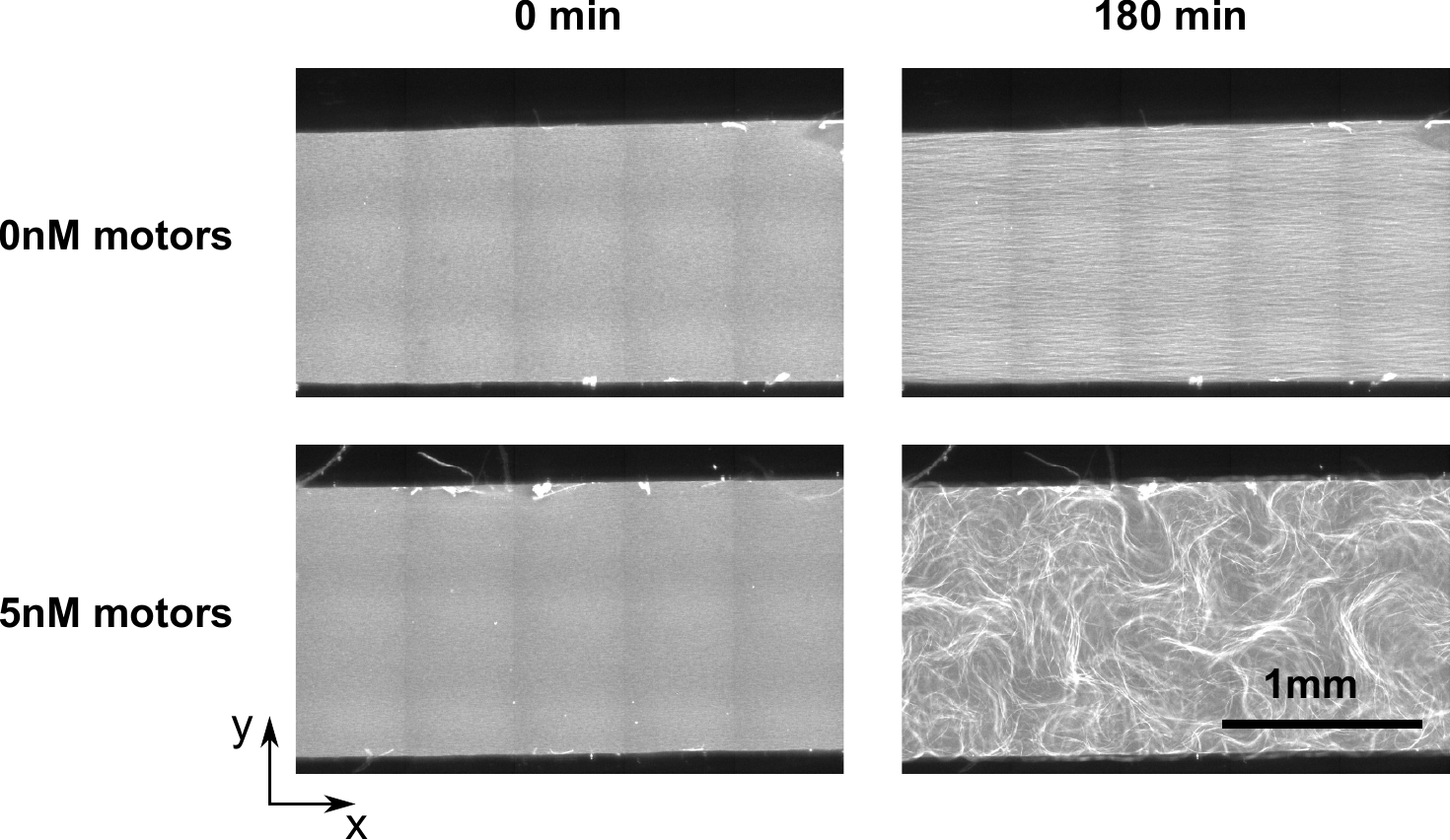}
		\renewcommand{\figurename}{Figure S\!\!}\caption{Epifluorescence images of an experiment performed with 1.5 $\mu$m-long GMPCPP-stabilized microtubules.
			No contraction of the gel is observed along $y$, neither in the presence nor in the absence of motors.
			Active turbulence is observed at 5 nM motors.}\label{Fig_SI_GMPCPP}
	\end{center}
\end{figure}

\begin{figure}[h!]
	\begin{center}
		\includegraphics[width=0.8\linewidth]{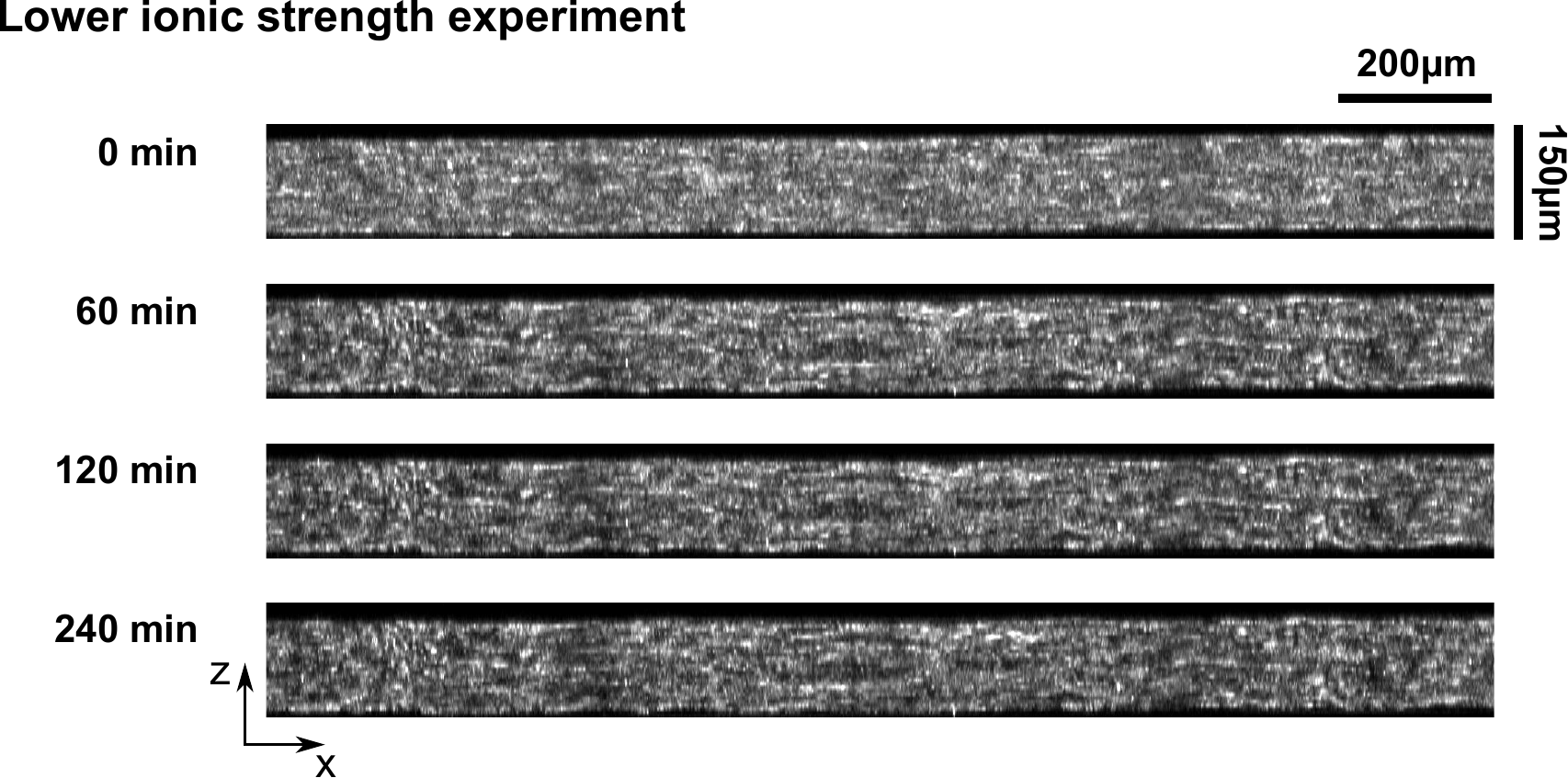}
		\renewcommand{\figurename}{Figure S\!\!}\caption{Reducing the attractive force between negatively charged, 8~$\mu$m-long, microtubules by lowering the ionic strength of the buffer impede the formation of the thin sheet and the buckling in the $xz$ plane. 0.2X PEM buffer, 2 mM K-acetate, 2 mM KCl, 1 mM MgCl$_{2}$ were used to prepare the active mix, with 0.5 nM motors, a $1/5$ dilution of these components compared to standard conditions.} \label{Fig_SI_less_ionic_strengt}
	\end{center}
\end{figure}

To the best of our knowledge, neither stable nor unstable out-of-plane buckling has been reported in active gels. We performed control experiments to determine which of the factors that differentiate our experiments from previously published 3D microtubule/kinesin gels\cite{Sanchez2012, Henkin2014, Wu2017} was responsible for the observed phenomenology: the type of motor or the length of the microtubules. We obtained both stable and unstable $xz$ buckling with the kinesin K401 used in previous reports\cite{Sanchez2012} (Figure S\ref{Fig_SI_K401BCCP}). This means that, although the motor K430 is not designed to form specific multimers, in contrast with K401, it forms non-specific ones. Our efforts to eliminate these non-specific multimers by size exclusion chromatography did not change the observed patterns (Figure S\ref{Fig_SI_akta}), suggesting that these clusters either form rapidly or do so in the working buffer. In contrast with the nature of the motor, the length of the microtubules had a strong impact on the observed structures. When, instead of 8~$\mu$m-long microtubules, 1.5~$\mu$m-long ones were used, no contraction of the gel was observed along $z$ or $y$, with or without motors, precluding the formation of a thin sheet that could buckle out of plane (Figure~S\ref{Fig_SI_GMPCPP}). In this case, the now classical active turbulence was observed at high activity, in agreement with previous reports\cite{Sanchez2012}. These observations are consistent with the expected linear dependence 
of the depletion free energy on filament length\cite{Braun2016} which, in our geometry, makes long microtubules condense into a thin sheet.
To further test this hypothesis we reduced the attractive force between negatively charged 8~$\mu$m-long microtubules by lowering the ionic strength of the buffer. In agreement with this interpretation, neither gel contraction, nor buckling in the $xz$ plane, was apparent when the salt concentration was diluted 5-fold (Figure~S\ref{Fig_SI_less_ionic_strengt}).

In summary, we demonstrate that in vitro active gels can be designed to form static or transient suspended sheets with periodic corrugated patterns of tunable wavelength. The mechanism of pattern formation that we propose combines passive and active processes that can be controlled physicochemically. Passive depletion forces, which depend on depletion agent concentration, filament length and ionic strength,  induce the spontaneous condensation of  a 3D nematic gel into a thin 2D nematic sheet, and active stresses buckle the gel sheet out of plane to form corrugations with well-defined wavelength that can be controlled by activity. In addition, we use an active gel theory to demonstrate that the observed patterns result from an out of plane  buckling instability induced by active extensile stresses along the nematic axis of the gel sheet, in contrast with in-plane buckling patterns that have been observed in pre-stressed nematic gels of either non-growing F-actin\cite{Gentry2009} and growing microtubules\cite{Liu2006} in the absence of motors. The buckling instability that we report does not involve filament flows and therefore fundamentally differs from both contractile instabilities in anisotropic active gels and spontaneous flow transitions in nematic active gels that have been described theoretically \cite{Bois2011, Kruse2004, Voituriez2005,  Edwards2009, Marchetti2013, Julicher2007}, and shown to be characterised by hydrodynamic flows and in-plane buckling of the director field in the case of 2D systems. Such spontaneous flows  have been observed in various active matter systems \cite{Sanchez2012, Wensink2012, Keber2014, Zhou2014, Kumar2014, DeCamp2015, Duclos2016, Duclos2018, Opathalage2019, Martinez-Prat2019}, which in practice  yield either turbulent or large scale coherent flows, but so far no static spatial patterns. 
In contrast, our results show that active matter can be shaped into long-lived static 3D patterns that can be tuned by activity, which may open the way to the design of 3D biomimetic materials\cite{Zadorin2017, Furuta2018}.

\section{Acknowledgements}
K. Furuta for providing the expression plasmids coding for the K430 kinesin, Z. Gueroui for a kind gift of the K401 plasmid, F. Lam from the microscopy platform at IBPS and L.L. Pontani for providing access to a spinning disk microscope, T. Surrey for insightful discussions and C. del Junco and Y. Vyborna for comments on the manuscript. This work has been funded by the European Research Council (ERC) under the European's Union Horizon 2020 programme (grant No 770940, A.E.-T.) and by the Ville de Paris Emergences programme (Morphoart, A.E.-T.)

\section{Supplementary materials}


\subsection{Methods\label{SI_methods}}

\subsubsection*{Chemicals and reagents}
All chemicals and reagents were purchased from Sigma-Aldrich, New England Biolabs, Roche, and ThermoScientific.

\subsubsection*{Kinesins expression}
From the plasmid coding for the K430 truncated kinesin-1 (amino acid residues 1-430) from Rattus norvegicus designed by Furuta et al. \cite{furuta2013}, we have built a homodimer version containing a SNAP-tag on each arm. Furuta's plasmid, pET-32ark430(C7S)-FlagC-rk430(C7S)-SnapC-His, which expresses both Kinesin-Snap-His and Kinesin-Flag was digested at two EcoRI sites to split into two fragments of Kinesin-Flag sequence and Kinesin-Snap-His with backbone sequence coding ampicillin resistance gene. The digested linear sequences were purified by QIAquick PCR Purification Kit (Qiagen) and recircularized by T4 DNA ligase using Rapid DNA Ligation Kit (Thermo Scientific). The recircularized plasmids were transformed into DH5$\alpha$ competent cells and selected on 100 $\mu$g/mL ampicillin including agar plate. The plasmids in colonies on agar plate were amplified in LB ampicillin medium and purified with Monarch Plasmid Miniprep Kit protocol (New England Biolabs). The sequence was verified by restriction enzyme digestion.

Homodimer K430 was expressed in competent cells, Rosetta2 (DE3) (Novagen). The cells were cultured in LB supplemented with 100 $\mu$g/mL Ampicilin and 34 $\mu$g/mL Chloramphenicol at 37 $^{\circ}$C until OD660 = 0.6. The protein expression was induced by 0.1 mM IPTG for 5 hours at 22 $^{\circ}$C. The cells were collected and resuspended in the Lysis Buffer: Buffer A (20 mM Na-Pi buffer pH 7.5, 1 mM MgSO$_{4}$, 250 mM NaCl, 0.015 mM ATP, 10 mM  $\beta$-mercaptoethanol, 0.1 \% Tween-20) supplemented with 10 mM imidazole and 1X protease inhibitor cocktail (Sigma-Aldrich). The cell suspension was then sonicated (VCX-130, Sonics Materials) and centrifuged. The supernatant was collected and filtrated before mixing with Ni-IMAC resin (Biorad). The resin was washed first with the Buffer A and 10 mM imidazole, and then with the same buffer but with 50 mM of imidazole. The resin was eluted with the Buffer A supplemented with 250 mM imidazole. The eluate was filtrated and filled in 14 kDa MWCO cellulose tube (Sigma-Aldrich) and then dialyzed with Buffer A three times (2 times 1 hour and overnight) at 4 $^{\circ}$C. The dialyzed protein was further purified with a Superdex 200 Increase column (GE Healthcare). The kinesin corresponding peak fraction was collected, flash-frozen and kept at $-80$~$^{\circ}$C.

For K401 purification, pT7-7\_DmKinesin 1-401 BCCP-CHis6 (pWC2) plasmid designed by Gelles J. J \cite{subramanian2007two} was expressed in competent cells Rosetta2 (DE3) (Novagen). The cells were cultured in LB supplemented with 100 $\mu$M biotin, 100 $\mu$g/mL ampicillin and 34 $\mu$g/mL chloramphenicol at 37 $^{\circ}$C until OD660 = 0.7. The protein expression was induced by 1 mM IPTG for 2 hours at 22 $^{\circ}$C and then the biotinylation was induced by 0.2 mM rifampicin for 20 hours at 22 $^{\circ}$C. The cells were collected and resuspended in the Buffer B (250 mM PIPES pH 7.2, 20 mM MgCl$_{2}$, 0.25 mM ATP, 50 mM $\beta$-mercaptoethanol) supplemented with 20 mM imidazole and 1X protease inhibitor cocktail (Sigma-Aldrich). The cell suspension was then sonicated (VCX-130, Sonics Materials) and centrifuged. The supernatant was collected and filtrated before mixing with Ni-IMAC resin (Biorad). The resin was washed with the Buffer B and 20 mM imidazole. The resin was eluted with the Buffer B and 500 mM imidazole. The eluate was filled in Float-A-Lyzer G2 (5 mL; 50K MWCO, Spectora/Por), dialyzed with Dialysis Buffer (250 mM PIPES pH 6.7, 20 mM MgCl$_{2}$, 0.25 mM ATP, 50 mM $\beta$-mercaptoethanol) three times (1 hour, 2.5 hours and overnight) at 4 $^{\circ}$C. The dialyzed protein supplemented with 36 \% sucrose and 2 mM dithiothreitol (DTT) was flash-frozen and kept at $-80$ $^{\circ}$C.

\subsubsection*{Microtubule polymerization}
Tubulin and TRITC-labeled tubulin were purchased from Cytoskeleton, dissolved at 10 mg/mL in 1X PEM buffer (80 mM PIPES pH 6.8, 1 mM EGTA, 1 mM MgSO$_{4}$) supplemented with 1 mM GTP, flash-frozen and stored at $-80~^{\circ}$C. 
The polymerization mix consists of 1X PEM, 1 mM GTP, 10 \% (w/v) glycerol and microtubules at 5 mg/mL (including 2.5 \% fluorescent tubulin). First the mix was centrifugated at 4 $^{\circ}$C for 15~min at 16000 g to remove small aggregates of tubulin. The corresponding supernatant was transferred into a new tube and incubated at 37 $^{\circ}$C for 15 min. 20 $\mu$M of paclitaxel (in the following taxol) was added to the mix and let at 37 $^{\circ}$C for five more minutes. After polymerization, newly formed microtubules were centrifugated at room temperature for 10 min at 12000 g to remove free tubulin monomers. The microtubules were redissolved into 1X PEM, 1 mM GTP, 10 \% glycerol, 20 $\mu$M taxol and kept in the dark at room temperature for few days.

For control experiments, GMPCPP (Jena Bioscience) microtubules were polymerized in the presence of 0.5 mg/mL GMPCPP from tubulin at 37 $^{\circ}$C for 30 min and left at room temperature for 5 hours. They were used within the same day.

\subsubsection*{Active mix}
The active mix consisted in 1X PEM buffer, 10 mM K-acetate, 10 mM KCl, 5 mM MgCl$_{2}$, 2 \%(w/v) Pluronic F-127, 5 $\mu$g/mL creatine kinase, 20 mM creatine phosphate, 20 $\mu$M taxol, 2 mM ATP, 1 mg/mL BSA, 1 mM trolox, 20 mM D-glucose, 3 mM DTT, 150 $\mu$g/mL glucose oxidase, 25 $\mu$g/mL catalase and 0.5 mg/mL taxol-stabilized microtubules.

\subsubsection*{Channel assembly}
Channels were assembled using a microscope glass slide (26 x 75 x 1 mm) and a coverslip (22 x 50 x 0.17 mm) separated by strips of Parafilm cut with a Graphtec Cutting Plotter CE6000-40. Both microscope glass slides and coverlips were passivated using an acrylamide brush \cite{sanchez2013}. The active mix was filled in the flow cell (22 x 1.5 x 0.130 mm) by capillarity and sealed with grease.

\subsubsection*{Imaging}
Epifluorescence images were obtained with a Zeiss Observer 7 automated microscope equipped with a Hamamatsu C9100-02 camera, a 10X objective, a motorized stage and controlled with MicroManager 1.4. Images were recorded automatically every 3 min using an excitation at 550 nm with a CoolLED pE2. Confocal images were obtained with a Leica TCS SP5 II confocal microscope with a 25x water-immersion objective or a X-Light V2 Spinning Disk Confocal system mounted on an upright Nikon Eclipse 80i microscope with a 10x objective. Images were recorded automatically every 1 to 10 min.

\subsubsection*{Image analysis}

Fluorescent images were binarized to obtain $\Delta\ell_z$ and  $\Delta\ell_y$. To measure $\phi$ the binarized $xz$ confocal cross-sections were averaged over $x$, smoothed along $x$ by applying a moving average filter with a 30-pixel window, that was then differentiated. $\phi$ was the arctangent of this derivative.

\newpage


\subsection{Hydrodynamic theory of an undulating active film\label{SI_theory}}

In this section, we discuss an active fluid model \cite{Marchetti2013, Ramaswamy2010, Prost2015, Julicher2007} to understand the buckling of the effectively two-dimensional sheet formed by the microtubules and obtain a theoretical estimate for the scaling of the wavelength of the undulatory pattern with motor concentration. The thickness of the nematic film, lying parallel to the the $xy$ plane, in the $z$ direction is denoted by $\ell_z$ and the channel thickness in the $z$ direction by $H$. The width of the channel in the $y$ direction is $W$ and the length along $x$ is $L$. The film is formed by injecting an isotropic microtubule fluid into the channel of dimensions $L\times W\times H$. The interaction mediated by depletion agents in the fluid then leads to the phase separation of the microtubule filaments and the fluid with a local concentration of filaments that exceeds the threshold for isotropic-nematic (I-N) transition. Due to the geometry of the channel, the nematic order develops along the long $x$ axis, leading to the formation of a film which contracts along the $y$ and $z$ direction and extends along the $x$ direction. This leads to a nematic film which is the thinnest along the $z$ direction. The orientation of the nematic is described by the director field $\hat{{\bf n}} = cos\:\phi \:\hat{x} + sin\: \phi\: \hat{z} = \hat{{\bf n}}_0 + \delta{\bf n}$, with the unbuckled state having the director $\hat{{\bf n}}_0=\hat{x}$. 

This film buckles along the $z$ direction due to the action of motors. We will now describe the physics of this buckling. We assume that the density of microtubules within the layer is fixed. We further assume that the nematic film buckles in the $z$ direction as whole. The departure of the mid-plane of the film from {a fiducial flat surface which we take to be the mid-$xy$ plane of the channel} is described by the height field {$h(x,y, t)$} {of the membrane point whose projection on that fiducial plane has the coordinates $(x,y)$}. Such a buckling is possible due to tilting of the microtubule filaments in the $xz$ plane $\delta{\bf n}\approx \partial_x h\hat{z}$ to linear order.

In equilibrium, tilting of filaments is penalised by the Frank elasticity \cite{deGennes1993}, considered here within a simplifying one Frank-constant approximation, which yields a bending modulus of the film:

\begin{equation}
\label{Frnk}
F_{\text{Frank}}=\frac{\bar{K}}{2}\int d^3{\bf r} (\nabla{\bf \hat{n}})^2\approx\frac{{K}}{2}\int d^2{\bf x}(\partial_x^2h)^2.
\end{equation}
where $K=\bar{K}\ell_z$ and ${\bf x}$ describes the $x-y$ plane. The membrane is confined in a channel of height $H$, which means that its mean-squared height fluctuations $\langle h^2\rangle\sim H^2$. Following \cite{Farago2008, Farago2004, Fournier2008}, we implement this constraint in an approximate manner by introducing a harmonic potential $(\gamma/2) h^2$ in the effective free energy of the membrane i.e.,
\begin{equation}
\label{frenrg}
F_h=\frac{1}{2}\int d^2{\bf x}\left[K(\partial_x^2 h)^2+\gamma h^2\right].
\end{equation}
In equilibrium at a temperature $T$, equating the mean-squared fluctuation with $H^2$ yields $\gamma=(K_BT/8H^2)^2/K$.
The membrane does not have a passive surface tension since we assume that it is free to adjust its area to maintain the preferred areal density of microtubules due to the depletion forces constant \cite{Safran2013}. However, in the experiments in which the channel is confined in the $x$ direction as well, the microtubule sheet may have a pattern even in the absence of motors due to Euler buckling. 
Since this is only manifested for channels confined along the $x$ axis and vanishes in other cases, we ignore this passive effect from here on and instead focus on the motor-driven periodic corrugation of the  active membrane.

We model the dynamics of an active membrane with the passive restoring forces arising from the derivative of a free-energy given by eq. \eqref{frenrg}. Since the membrane is suspended in a fluid, it moves due to the motion of the fluid. Since $h$ represent the height of a membrane point above a fiducial surface, $\dot{h}\sim v_z|_m$ where $v_z|_m$ is the $z$ component of the fluid velocity at the membrane position. For the linear theory that we are interested in here, $v_z|_m$ can be approximated by $v_z(0)$ the $z$ component of the velocity at the position of the fiducial plane. However, if fluid can pass (permeate) through the membrane, fluid flow need not correspond to the motion of the membrane. This relative velocity of the fluid and a passive membrane is $\propto \mu\delta F_h/\delta h$, where $\mu$ is the permeation coefficient. In addition, active membranes may move relative to the background fluid due to an active speed $\sim -\mu \bar{\zeta}(c)\hat{{\bf n}}_0\cdot\nabla\delta{\bf n}\approx -\mu \bar{\zeta}(c)\partial_x^2 h\hat{z}$ \cite{Maitra2014}. For extensile filaments, the active coefficient, whose magnitude depends on the motor concentration, $\bar{\zeta}(c)>0$. Thus, the hydrodynamic equation for $h$ is 

\begin{equation}
	\label{memeq}
	\dot{h}-v_z(0)=-\mu\frac{\delta F_h}{\delta h}-\mu \bar{\zeta}(c)\partial_x^2 h
\end{equation}

The Stokes equation for the velocity field containing the membrane composed of active units is 
\begin{equation}
	\label{vel}
	\eta\nabla^2{\bf v}=\nabla\Pi+\zeta(c)\nabla\cdot({\bf nn})\delta(z)+\frac{\delta F_h}{\delta h}\hat{z}\delta(z)\approx\nabla\Pi+\left(\zeta(c)\partial_x^2 h+\frac{\delta F_h}{\delta h}\right)\hat{z}\delta(z)+\boldsymbol{\xi}_v
\end{equation}
where  $\Pi$ is the pressure that enforces the incompressibility constraint $\nabla\cdot{\bf v}=0$, $\zeta$ is the coefficient of the standard active stress $\zeta(c)>0$ for extensile systems. Note that, in principle, $\zeta(c)$ and $\bar{\zeta}(c)$ can be different from each other, though their dependence on motor concentration is expected to be the same. To obtain the second approximate equality, we have ignored all fluctuations of the nematic director in the $xy$ plane (we will discuss the consequence of this approximation later).

The fluid is in a channel of thickness $H$. We now {use the Fourier transformed version of eq. \eqref{vel} to} calculate $v_z(0)$ by summing {over all Fourier modes for which} $|q_z|>2\pi/H$, which takes the constraint due to the channel into account in an approximate manner. Eliminating the pressure by projecting the Fourier transformed velocity transverse to the wavevector direction, we obtain
{\begin{equation}
		v_z(0)=\frac{-1}{2\pi\eta}\left[\int_{\frac{2\pi}{H}}^\infty dq_z\frac{q_x^2}{(q_x^2+q_z^2)^2}G_{z}(q_x)+\int_{-\infty}^{-\frac{2\pi}{H}} dq_z\frac{q_x^2}{(q_x^2+q_z^2)^2}G_{z}(q_x)\right]
\end{equation}}
where $G_z(q_x)$ is the Fourier transform of $\zeta(c)\partial_x^2 h+{\delta F_h}/{\delta h}$. This yields
{\begin{equation}
		v_z(0)=-\frac{1}{4\eta|q_x|}\left[1-\frac{4H|q_x|}{4\pi^2+H^2q_x^2}-\frac{2}{\pi}\tan^{-1}\left(\frac{2\pi}{H|q_x|}\right)\right]G(q_x)
\end{equation}}
The effective mobility goes to $-1/4\eta|q_x|$ in the limit $H\to\infty$ \cite{Brochard1975} and $\sim q_x^2H^3$ for $q_xH\ll 1$ \cite{Gov2004, Seifert1994}.
The force 
{\begin{equation}
		G(q_x)=(Kq_x^4-\zeta(c)q_x^2+\gamma)h_{\bf q}.
\end{equation}}
Putting this back in the equation for height fluctuations, and Fourier transforming in space and time, we obtain
{\begin{equation}
		\omega=-\frac{i}{4\eta|q_x|}\left[1-\frac{4H|q_x|}{4\pi^2+H^2q_x^2}-\frac{2}{\pi}\tan^{-1}\left(\frac{2\pi}{H|q_x|}\right)\right](Kq_x^4-\zeta(c)q_x^2+\gamma)-i\mu(Kq_x^4-\bar{\zeta}(c)q_x^2+\gamma)
\end{equation}}
Since both the hydrodynamic mobility and the permeation coefficient $\mu$ are always positive, we clearly see that extensile activity leads to a positive growth rate for a band of wavenumbers. If permeation dominates over hydrodynamics, a band of wavevectors between 
\begin{equation}
	\label{unstab}
	q_{x_\pm}^2=\frac{\bar{\zeta}_c\pm\sqrt{\bar{\zeta}^2-4K\gamma}}{2}
\end{equation}
is unstable. Note that, unlike usual active planar instabilities, the instability here is not long wavelength i.e., the membrane is stable for $q_x\to 0$ due to the confinement. The fastest growing mode is 
\begin{equation}
	q^*_x=\sqrt{\frac{\bar{\zeta}(c)}{2K}}.
\end{equation}
We identify this with the wavevector of the pattern we observe. Further, it is known \cite{Martinez-Prat2019} that active forces scale as $c^2$. This implies that the wavelength of the buckled pattern should scale linearly with $c$, i.e., $\lambda=2\pi/q^*\propto c$. This linear scaling is observed in our experiments.

If hydrodynamics dominates over permeation (such that permeation can be ignored), the unstable band of wavevectors is given by \eqref{unstab}, but with $\bar{\zeta}(c)$ replaced by $\zeta(c)$. Furthermore, in the limits  $Hq_x\to\infty$ and $Hq_x\to 0$, the fastest growing mode is 
\begin{equation}
	q^*_x=\sqrt{\frac{{\zeta}(c)}{2K}}.
\end{equation}
While the observed selected wavelength in our system is in neither regime, $H q_x=2\pi H/\lambda\sim \mathcal{O}(1)$, we do not believe that this will significantly change the scaling with ${\zeta}$. Further, since $\zeta(c)$ and $\bar{\zeta}(c)$ should scale the same way with $c$, the scaling of the wavelength of the pattern with $c$ remains the same, $\lambda\propto c$ for both permeation dominated and hydrodynamics dominated cases. However, the hydrodynamics dominated and the permeation dominated cases can be distinguished by looking at the dependence of the characteristic frequency on activity. For the permeation dominated case,
\begin{equation}
	i\omega^*\sim \mu(Kq_x^{*^4}-\bar{\zeta}(c)q_x^{*^2}+\gamma).
\end{equation}
This implies that $i\omega^*\sim \bar{\zeta}^2\sim c^4$. For the hydrodynamics dominated case, and for $q_xH\gg 1$, $i\omega^*\sim {\zeta}^{3/2}\sim c^3$ while for $q_xH\ll 1$, $\omega^*\sim {\zeta}^{3}\sim c^6$.

We have primarily discussed the fluctuation of the thin film along the $z$ direction. The film also contracts along the $y$ direction. A similar theory could be constructed for the undulations of the film in the $y$ direction as well. Since the film is thicker in this direction, thus having a larger bending modulus, $K_w\sim \bar{K}w$, where $w$ is the width of the film, these undulations have a longer wavelength since $q_w^*\sim1/\sqrt{K_w}$ and $K_w>K\implies q^*_w<q^*$. However, when the anisotropy of the channel is reduced such that $W\sim H$ and therefore, $w\sim \ell_z$, buckling in both $y$ and $z$ direction should be observed simultaneously and should have roughly equal wavelengths. 
Ultimately, at high activities the coupling of undulations in $y$ and $z$ directions leads to the destruction of the periodically modulated state. 
Finally, We have not considered the fluctuation of the nematic order of the microtubules in the plane of the membrane. Taking into account such fluctuations is complicated (see \cite{Nelson1987, Nelson1993}) and, in any case, would not change the linear physics of the membrane modulation discussed here.

\newpage


\subsection{Supplementary figures}

\newpage

\newpage

\newpage

\newpage

\newpage

\newpage

\newpage

\newpage

\newpage

\newpage

%



\newpage

\subsection{Supplementary videos}
\setcounter{figure}{0}

\begin{figure}[h!]
	\centering
	\includegraphics[width=0.5\linewidth]{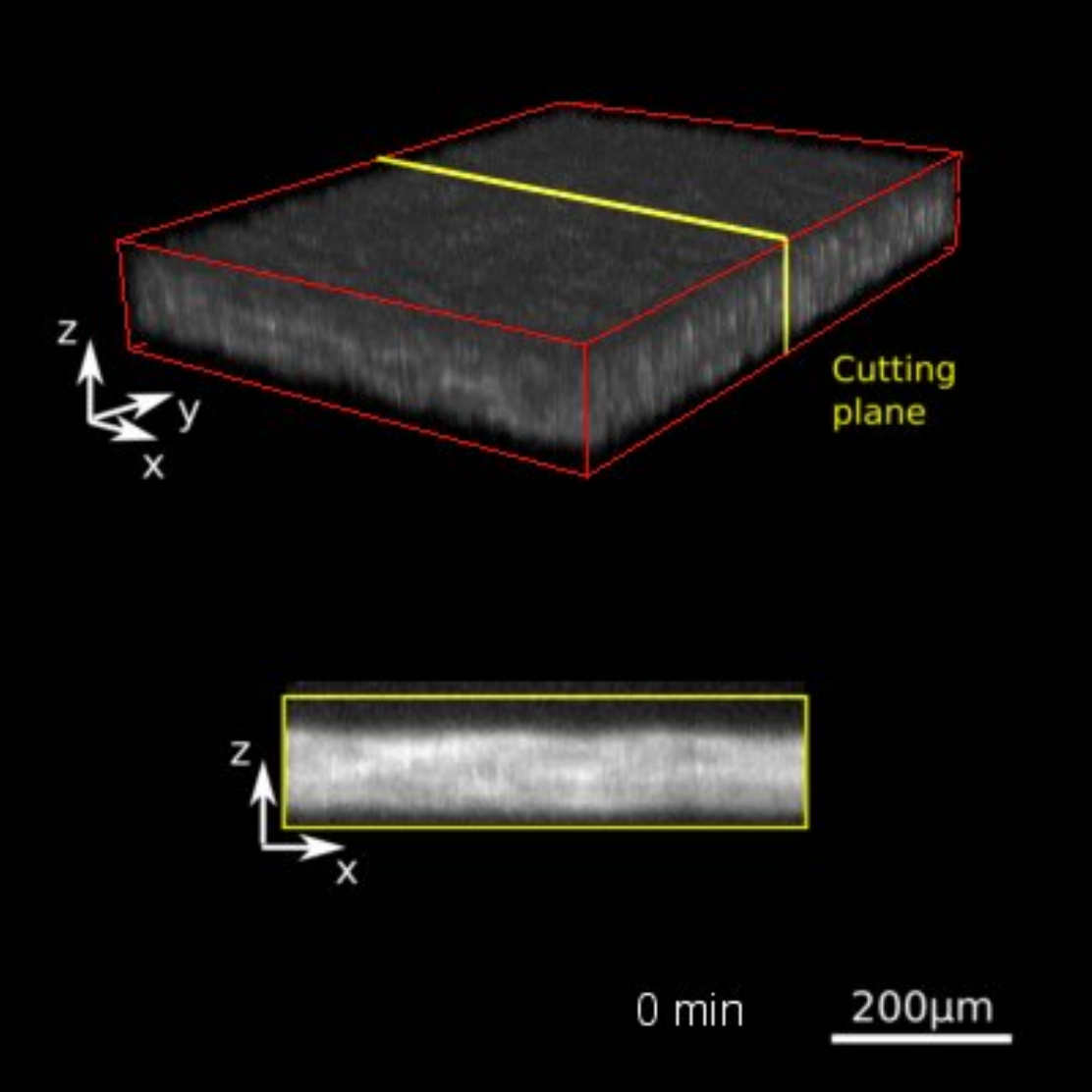}
	\renewcommand{\figurename}{Video S\!\!}\caption{Confocal images and $xz$ sectioning of a 3D extensile, nematic gel made of microtubules and kinesin motors forming a corrugated sheet. Motor concentration is 0.5 nM.}\label{Video_SI_video_fig2MT} 
\end{figure}

\newpage

\begin{figure}[h!]
	\centering
	\includegraphics[width=1\linewidth]{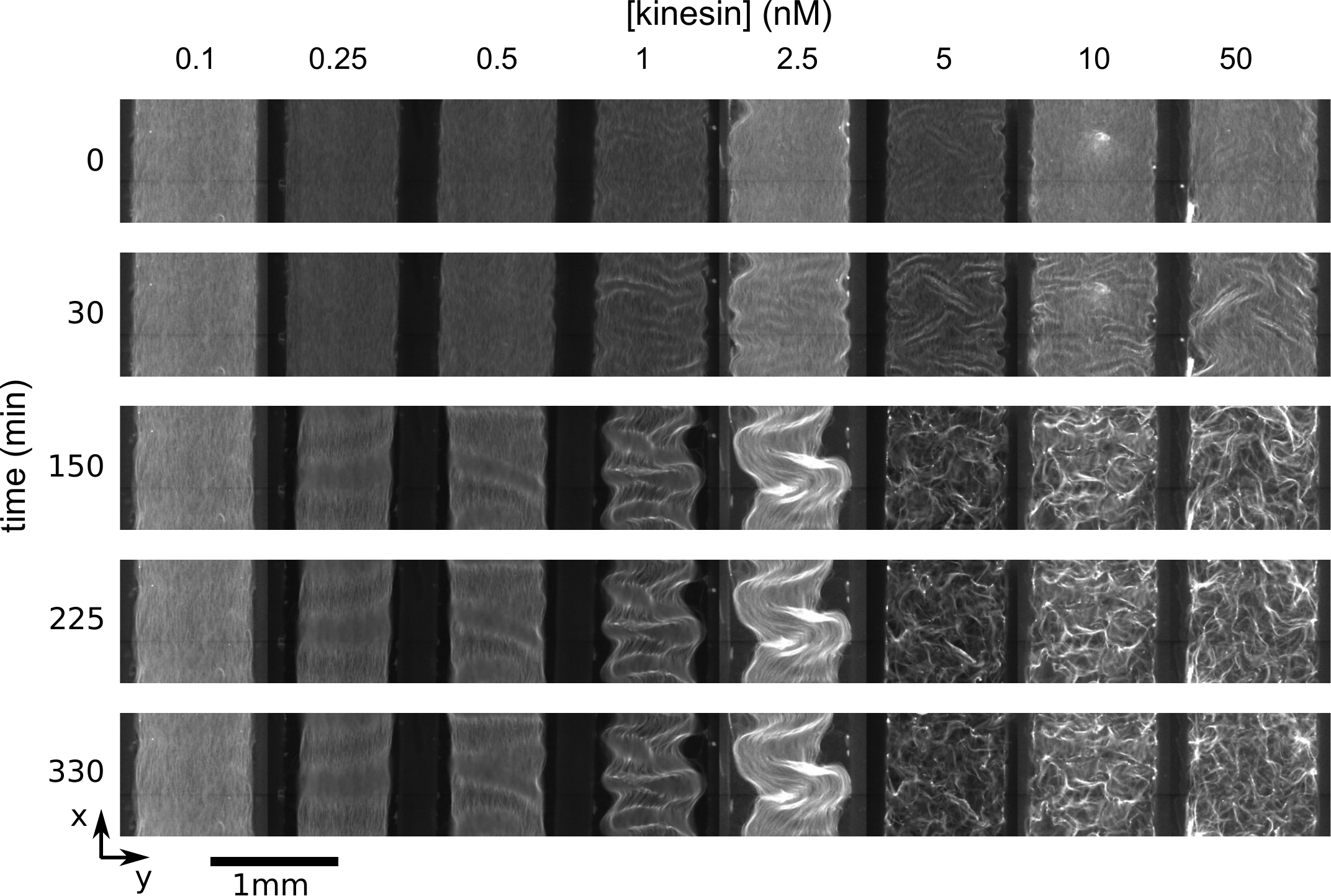}
	\renewcommand{\figurename}{Video S\!\!}\caption{
		Dynamics, shape and stability of the patterns depending on the motor concentration. The video shows different channels with different motor concentrations recorded in epifluorescence.}\label{Video_SI_epifluo_kin_range} 
\end{figure}

\newpage

\begin{figure}[h!p]
	\centering
	\includegraphics[width=0.9\linewidth]{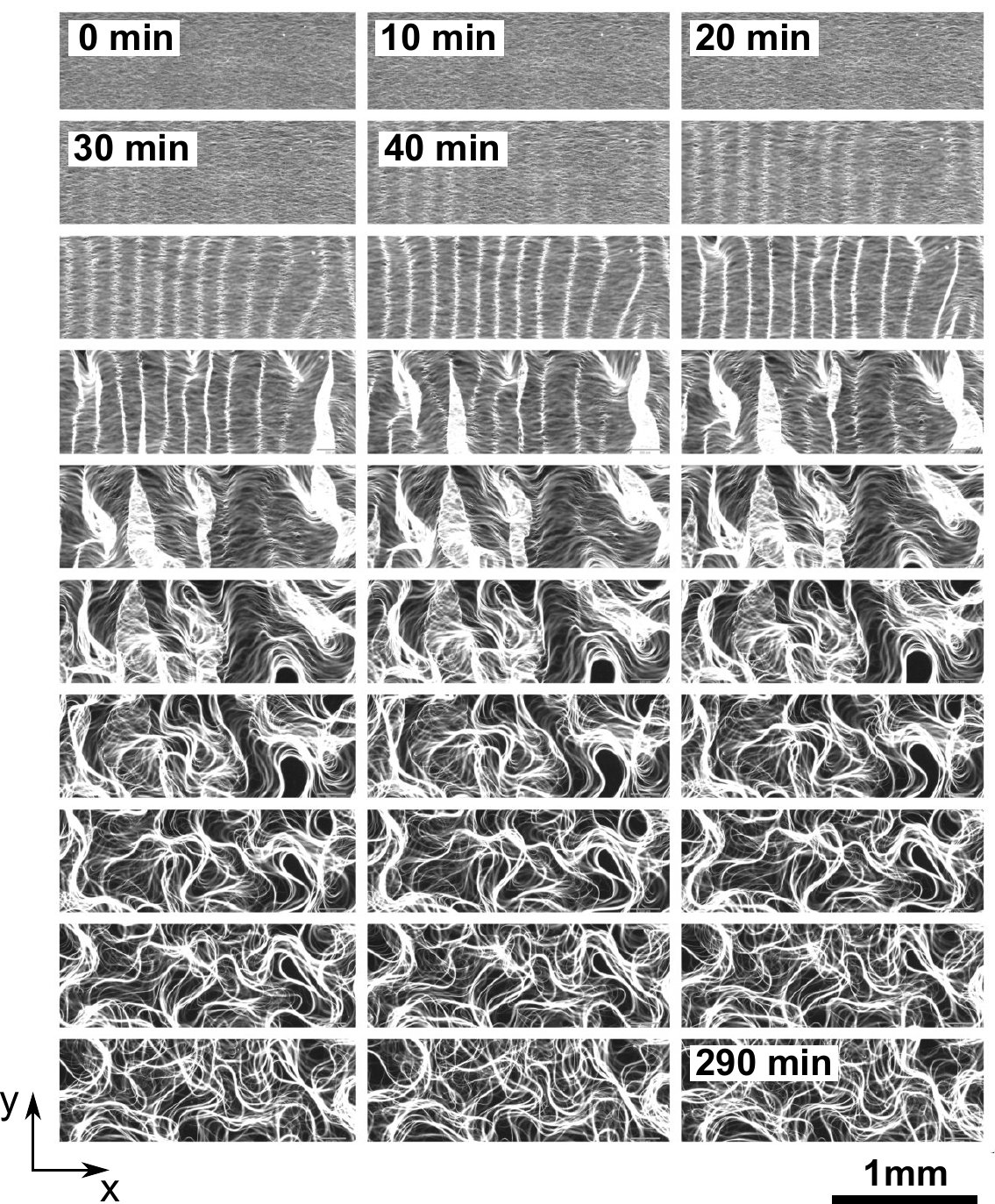}
	\renewcommand{\figurename}{Video S\!\!}\caption{A 3D extensile, nematic gel made of microtubules and kinesin motors forms a corrugated sheet that breaks into active turbulence at high motor concentration. Buckling in the $xz$ plane is observed at early times (stripes at 60 min) then the pattern breaks into an active turbulent state. Epifluorescence time-lapse images in the $xy$ direction. The movie is 10 hours long. This experiment corresponds to a 0.75 nM motor concentration of K430 before size exclusion chromatography and thus its activity is stronger than the rest of the experiments, where purified K430 was used.} \label{Video_SI_video_BandsToTurbulence} 
\end{figure}

\newpage

\begin{figure}[h!]
	\centering
	\includegraphics[width=0.75\linewidth]{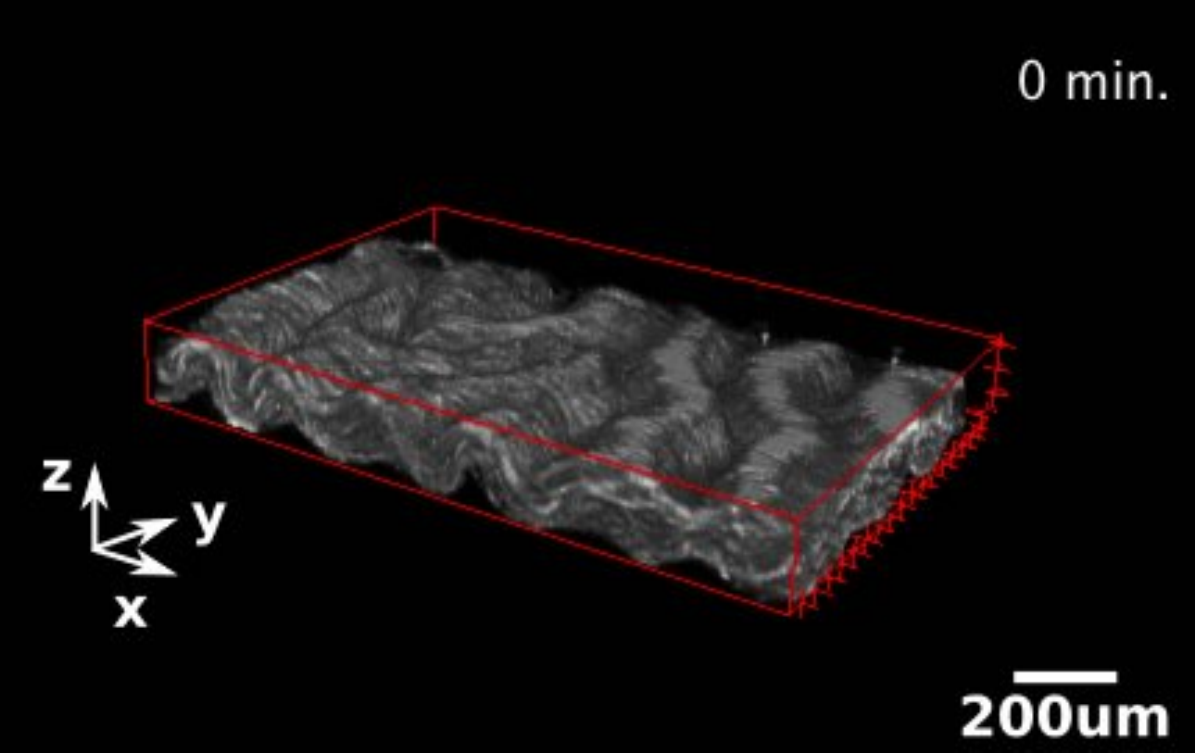}
	\renewcommand{\figurename}{Video S\!\!}\caption{Confocal images of a 3D extensile, nematic gel made of microtubules and kinesin motors forming a corrugated sheet that breaks into active turbulence. Motor concentration is 5 nM. The formation of the corrugated sheet from the initial nematic state is not visible here as it occurs at the very beginning, during the sample preparation.}\label{Video_SI_confocal_bands_to_turbulence} 
\end{figure}

\newpage



\providecommand{\latin}[1]{#1}
\makeatletter
\providecommand{\doi}
  {\begingroup\let\do\@makeother\dospecials
  \catcode`\{=1 \catcode`\}=2\doi@aux}
\providecommand{\doi@aux}[1]{\endgroup\texttt{#1}}
\makeatother
\providecommand*\mcitethebibliography{\thebibliography}
\csname @ifundefined\endcsname{endmcitethebibliography}
  {\let\endmcitethebibliography\endthebibliography}{}



\end{document}